\documentclass[%
 reprint,
 amsmath,amssymb,aps,
longbibliography]{revtex4-2}

\usepackage{graphicx}
\usepackage{dcolumn}
\usepackage{bm}

\usepackage{enumitem}
\usepackage{soul}
\usepackage{ulem}
\usepackage{xcolor}
\usepackage{multirow}
\usepackage{url}

\usepackage{amsmath}
\usepackage{amsthm}
\usepackage{amssymb}
\usepackage{amsfonts}
\usepackage{enumerate}
\usepackage{braket}
\usepackage[utf8]{inputenc}
\usepackage[top=1 in,bottom=1in, left=1 in, right=1 in]{geometry}
\usepackage{tikz}
\usetikzlibrary{quantikz2}
\usepackage{tabularx}

\begin{document}

\preprint{APS}

\title{Engineering Non-Gaussian Bosonic Gates through Quantum Signal Processing}

\author{Pak-Tik Fong}
\email{Email: ptf@sfu.ca}
\author{Hoi-Kwan Lau}
\affiliation{%
Department of Physics, Simon Fraser University, Burnaby, British Columbia, Canada, V5A 1S6}


\date{\today}

\begin{abstract}

Non-Gaussian operations are essential for most bosonic quantum technologies. Yet, realizable non-Gaussian gates are rather limited in type and generally suffer from accuracy-duration trade-offs. In this work, we propose to use quantum signal processing (QSP) techniques to engineer non-Gaussian gates on hybrid qumode-qubit systems. For systems with dispersive coupling, our scheme can generate a new non-Gaussian gate that produces a phase shift depending on the modulus of the boson number. This gate reproduces the selective number-dependent arbitrary phase (SNAP) gates under certain parameter choices, but with higher accuracy within a short, fixed and excitation-independent interaction time. The gate unlocks new applications, for example, in entangling logical qudits and deterministically generating multi-component cat states. Additionally, our versatile QSP formalism can be extended to systems with other interactions, and also engineer non-unitary operations, such as noiseless linear amplification and generalized-parity measurement. 

\end{abstract}

\maketitle


\section{Introduction}

In bosonic quantum systems \cite{Kimble1998, RevModPhys.93.025005, RevModPhys.86.1391,PhysRevLett.103.070502, RevModPhys.82.1041}, Gaussian gates constitute the most commonly implemented class of unitary operations \cite{rmp2005,rmp2012, fong2024}. These are the operations that transform a Gaussian state only to another Gaussian state \cite{WANG20071}. Despite their limitations, they enable diverse quantum technological applications, including linear amplification \cite{caves1982}, secure quantum communication \cite{PhysRevA.61.010303, PhysRevLett.88.057902, PhysRevA.65.042310,PhysRevA.88.042313}, quantum sensing \cite{cv_sensing2000, su11, metrology2010, singlemode2017} and many more. Nevertheless, it has been known that if only Gaussian state preparation and measurement are available, Gaussian gates alone are not sufficient to implement various tasks, such as universal continuous-variable (CV) quantum computing (QC) \cite{Lloyd1999}, error correction \cite{CAI202150, nogo_QEC} and entanglement distillation \cite{ impossible_distill2002}.

Non-Gaussian gates are therefore valuable resources for unlocking the full potential of bosonic quantum technologies \cite{rmp2005}. One way to generate non-Gaussian gates is to apply an interaction that involves higher-than-quadratic order in mode operators. However, these interactions are usually weak in bosonic platforms, such as optics \cite{PhysRevA.84.053802, PhysRevLett.125.160501, PhysRevResearch.6.023332, PhysRevA.97.022329}. An alternative way is to engineer hybrid quantum systems in which the bosonic mode (qumode) is strongly coupled to an ancillary qubit \cite{Andersen2015, liu2024}, such as via dispersive \cite{Schuster2007} or Jaynes-Cummings (JC) couplings \cite{JC_1963}. Nevertheless, directly employing these couplings typically entangles the qumode with the qubit; this potentially disturbs the encoded bosonic information. Consequently, implementing a non-Gaussian gate requires carefully controlling the hybrid interaction to disentangle the qumode and qubit after the operations. 

One approach involves the use of multi-tone driving, a technique frequently employed in circuit quantum electrodynamics (QED) platforms \cite{PRXQuantum.3.030301, PhysRevLett.133.260802}. For example, selective number-dependent arbitrary phase (SNAP) gates \cite{Schuster2007, SNAP_2015, SNAP2015_2} can be engineered under dispersive coupling and multi-tone driving.
Taking advantage of the uneven dressed-state energy splitting induced by dispersive coupling, each bosonic level can be individually manipulated. By carefully choosing the magnitude and phase of each tone, the desired phase shift can be exerted while the qumode and qubit remain separable after the interaction. However, this individual manipulability requires the driving tones to be weak to suppress the off-resonant effects, and hence it will limit the gate speed. 

Alternatively, one can implement non-Gaussian gates by adiabatically tuning the interaction. By utilizing the non-linearity of hybrid interactions and the fact that each level remains an instantaneous eigenstate during adiabatic evolution \cite{PhysRevA.49.R3174}, transformation beyond Gaussian gates can be implemented. By choosing the hybrid interaction to be far off-resonant at the beginning and the end of the process, this can guarantee that the qubit will be disentangled after the operations. This method has been applied in trapped-ion systems to demonstrate excitation addition and subtraction \cite{Um2016}. However, maintaining adiabaticity generally leads to slower operations.

These methods typically involve a trade-off between the gate's accuracy and its speed. Various attempts have recently been proposed to mitigate this trade-off, for example, by optimizing the time-dependence of the drives \cite{PRXQuantum.3.030301, PhysRevLett.133.260802}, shortcuts to adiabaticity \cite{Demirplak2003, 10.1063/1.2992152, Berry_2009, PhysRevA.100.032323} or time-varying interaction amplitudes \cite{ FigueiredoRoque2021}. Nevertheless, these techniques often require numerical optimizations which obscure the underlying physics and whose complexity grows dramatically with the number of involved levels.

In this work, we introduce a novel method to engineer non-Gaussian operations by using the recently developed techniques of quantum signal processing (QSP) \cite{QSP2016, QSP2017, QSP2019}. QSP originates from the composite-pulse techniques for spin control \cite{LEVITT198661, Odedra2011}. By applying a well-designed sequence of drives, the spin evolution can be engineered to arbitrarily depend on its energy splitting. In hybrid qumode-qubit systems, the hybrid coupling induces an auxiliary qubit energy splitting that depends on the boson number of the qumode. Thus, the evolution of each boson-number state can also be engineered individually via QSP.
We will show that our QSP method is applicable to engineer gates with a variety of hybrid couplings, including the commonly found dispersive and JC couplings, and can generally reduce implementation time without compromising accuracy. 

Moreover, under dispersive coupling, QSP can engineer a new type of non-Gaussian gate, the mod-$k$ gate. It is a generalization of the exponential-Parity (eParity) gate which, together with beam-splitters, can be used as logic gates for any bosonic code \cite{ep_gate_2017, Gao2019} or highly mixed states \cite{PhysRevA.95.022303,PhysRevA.99.032345}. Remarkably, the implementation time of the mod-$k$ gates involves only two periods of dispersive coupling and is independent of the number of qumode levels involved.
Our new gate opens up new utilities, such as entangling qudits, manipulating bosonic codes and deterministically creating multi-component cat states.

Our QSP framework can also be extended to engineer non-unitary operations for, e.g., noiseless linear amplification (NLA) with optimal success rate \cite{NLA_2009, NLA_limit_2013} and generalized-parity measurement. 
We note that our method differs from the recent attempts to apply QSP in CV quantum information processing (QIP) \cite{liu2024, QSP_liu2024, crane2024, Park2024, zeytinoglu2024}, which focus on approximately simulating quantum evolution. Our approach can engineer the exact quantum gates and measurements under a finite implementation time. 

The paper is structured as follows. In Sec.~\ref{sec_QSP}, we review the QSP framework and outline its application in engineering non-Gaussian gates. Sec.~\ref{sec_mod_k} introduces mod-$k$ gates and their construction in dispersive coupling systems. Some applications of mod-$k$ gates are discussed in Sec.~\ref{sec_app}. The realization of QSP in JC systems is discussed in Sec.~\ref{sec_JC}. The extension of our method for non-unitary operations is discussed in Sec.~\ref{sec_non_unitary}. Finally, we conclude our findings in Sec.~\ref{sec_conclusion}.

\section{Quantum signal processing}\label{sec_QSP} 

QSP is a generalized composite-pulse technique, which aims at tailoring the transformation of a two-level system \cite{QSP2016, QSP2017, QSP2019, QSP2021, GQSP2024}. 
It is composed of a sequence of alternative phase-shifts, $\hat{Z}_\Phi$, and rotations, $\hat{R}_m$, as depicted in Fig.~\ref{QSP_circuit}(a). In this framework, every phase-shift operation shifts the same phase angle $\Phi$; its expression in the computational basis $\{\ket{\downarrow} \equiv (1,0)^T, \ket{\uparrow}\equiv (0, 1)^T\}$ is given by
\begin{eqnarray}
 \hat{Z}_\Phi &\equiv& 
 \begin{pmatrix}
e^{i\Phi} & 0 \\
0 & 1
\end{pmatrix}. \label{Z_phi} 
\end{eqnarray}
On the other hand, the angles are generally distinct for each rotation. We define the $m$-th round of rotation as
\begin{eqnarray}     \label{rotation_matrix} 
\hat{R}_m \equiv 
 \begin{pmatrix}
e^{i(\lambda_m + \phi_m)}\cos \theta_m & e^{i\phi_m}\sin\theta_m \\
e^{i\lambda_m}\sin\theta_m & -\cos\theta_m
\end{pmatrix}, \label{R}
\end{eqnarray}
where $\theta_m,\phi_m,\lambda_m \in [0,2\pi)$ are the corresponding rotation angles. We note that QSP is sometimes formulated as a method to engineer the transformation of a multi-level system through an auxiliary control qubit \cite{GQSP2024}. This formalism is equivalent to our model if one considers the eigenbasis of the multi-level system, which the joint system-ancilla can be represented by a collection of independent two-level systems. More details are provided in Appendix~\ref{appendix_big_matrix}.

\subsection{Original QSP} 

The original formulation of QSP (oQSP) \cite{QSP2016, QSP2017, QSP2019} utilizes the rotations along only a single axis, i.e. when all $\phi_m = \lambda_m = \pi/2$ but arbitrary $\theta_m$. This formulation is particularly useful for the situations where the rotation axis is restricted, such as in JC systems, which will be discussed in Sec.~\ref{sec_JC}.
After a sequence of $M$ rounds of phase-shifts and rotations, the overall operation becomes
\begin{eqnarray}\label{QSP_sequence_1}
    \hat{U}_{\text{QSP}} = \prod_{m=1}^{M} \left(\hat{R}_{m}\hat{Z}_\Phi\right) \hat{R}_0 = \begin{pmatrix}
F(e^{i\Phi}) & G(e^{-i\Phi}) \\
G(e^{i\Phi}) & F(e^{-i\Phi})
\end{pmatrix}.\nonumber\\
\end{eqnarray}
Here $F(z)$ and $G(z)$ are degree-$M$ polynomials of a complex variable $z$:
\begin{eqnarray}
    F(z) =  \sum_{m=0}^M f_m z^{m} \quad \text{and} \quad
    G(z) =  \sum_{m=0}^M ig_m z^{m},\label{G_z}
\end{eqnarray}
where the coefficients $f_m$ and $g_m$ are real. 
By varying the rotation angles $\{\theta_m\}$, these coefficients can be engineered to take arbitrary real values that satisfy the relation
\begin{eqnarray} \label{FG_constraint}
    |F(z)|^2+|G(z)|^2 = 1,
\end{eqnarray}
for all $z$ on the unit circle, i.e. $|z|=1$. This relation originates from the conservation of the total two-level system state population. We note that the expression and structure of the polynomials depend on the conventions adopted in the literature. For example, in Refs.~\cite{QSP2016, QSP2017} the polynomials are functions of $\cos(\Phi/2)$, while in Ref.~\cite{QSP2019} they appear as Laurent series in $e^{i\Phi/2}$. In this work, we adopt a variant of the latter convention by expressing the polynomial in terms of $z\equiv e^{i\Phi}$; this choice aligns naturally with the formalism of the Generalized QSP (GQSP) that will be described in the next subsection. The relations between different QSP conventions are discussed in Appendix.~\ref{appendix_convention}.

The goal of QSP is to design the polynomials so that the two-level system undergoes the desired transformation. In many applications, only the $\ket{\downarrow}$ component of the evolved state is considered, so we need to design only the polynomial $F(z)$, which should satisfy 
\begin{eqnarray}\label{constraint_of_F}
    |F(z)|\leq 1.
\end{eqnarray}
Once $F(z)$ is designed, the other polynomial $G(z)$ can always be determined up to a phase factor \cite{QSP2016}. After the coefficients of both polynomials are specified, the required rotation angles can be calculated by, for example, the algorithm described in Ref.~\cite{chao2020findinganglesquantumsignal}.

\subsection{Generalized QSP}\label{gqsp_section}

When rotations along arbitrary axes are available, i.e. $\lambda_m, \phi_m$ and $\theta_m$ are all controllable, GQSP can be realized \cite{GQSP2024}. This relaxes some constraints on the attainable polynomials. Specifically, after $M$ rounds of phase-shifts and rotations, the overall transformation is given by
\begin{eqnarray}\label{QSP_sequence_2}
    \hat{U}_{\text{GQSP}}  = \prod_{m=1}^{M} \left(\hat{R}_{m}\hat{Z}_\Phi\right) \hat{R}_0 =  \begin{pmatrix}
P(e^{i\Phi}) & \boldsymbol{\cdot} \\
Q(e^{i\Phi}) & \boldsymbol{\cdot}
\end{pmatrix}.
\end{eqnarray}
Since our protocols always consider the system that is initially in $\ket{\downarrow}$, we omitted the second column of the matrix~\eqref{QSP_sequence_2}, which corresponds to the transformation amplitudes if the system is initially $\ket{\uparrow}$.  

In this framework, the transformation amplitudes are determined by the following degree-$M$ polynomials,
\begin{eqnarray}\label{eq_of_p_q}
    P(z) =  \sum_{m=0}^M p_m z^{m} \quad \text{and} \quad Q(z) = \sum_{m=0}^M q_m z^{m}.
\end{eqnarray}
The coefficients $p_m$ and $q_m$ can be engineered to any complex values that satisfy the constraint
\begin{eqnarray}\label{PQ_constraint}
    |P(z)|^2+|Q(z)|^2 = 1,
\end{eqnarray} 
for every $z$ on the unit circle. Compared to the real-coefficient polynomials by oQSP (cf. Eq.~\eqref{G_z}), the complex-coefficient polynomials permitted by GQSP generally provide more freedom to match the needs of the applications. This can lower the required degree of the polynomials being engineered, and hence reduce the engineering complexity.

Analogously to oQSP, the existence of a $Q(z)$ is guaranteed if $P(z)$ satisfies the condition
\begin{eqnarray}\label{constraint_of_P}
    |P(z)|\leq 1.
\end{eqnarray}
Once the desired $P(z)$ and $Q(z)$ are determined, the corresponding rotation angles can be systematically determined \cite{GQSP2024}.

Unless otherwise specified, the term ``QSP" hereafter will collectively refer to both oQSP and GQSP. 

\begin{figure*}
    \centering
\includegraphics[width=\textwidth]{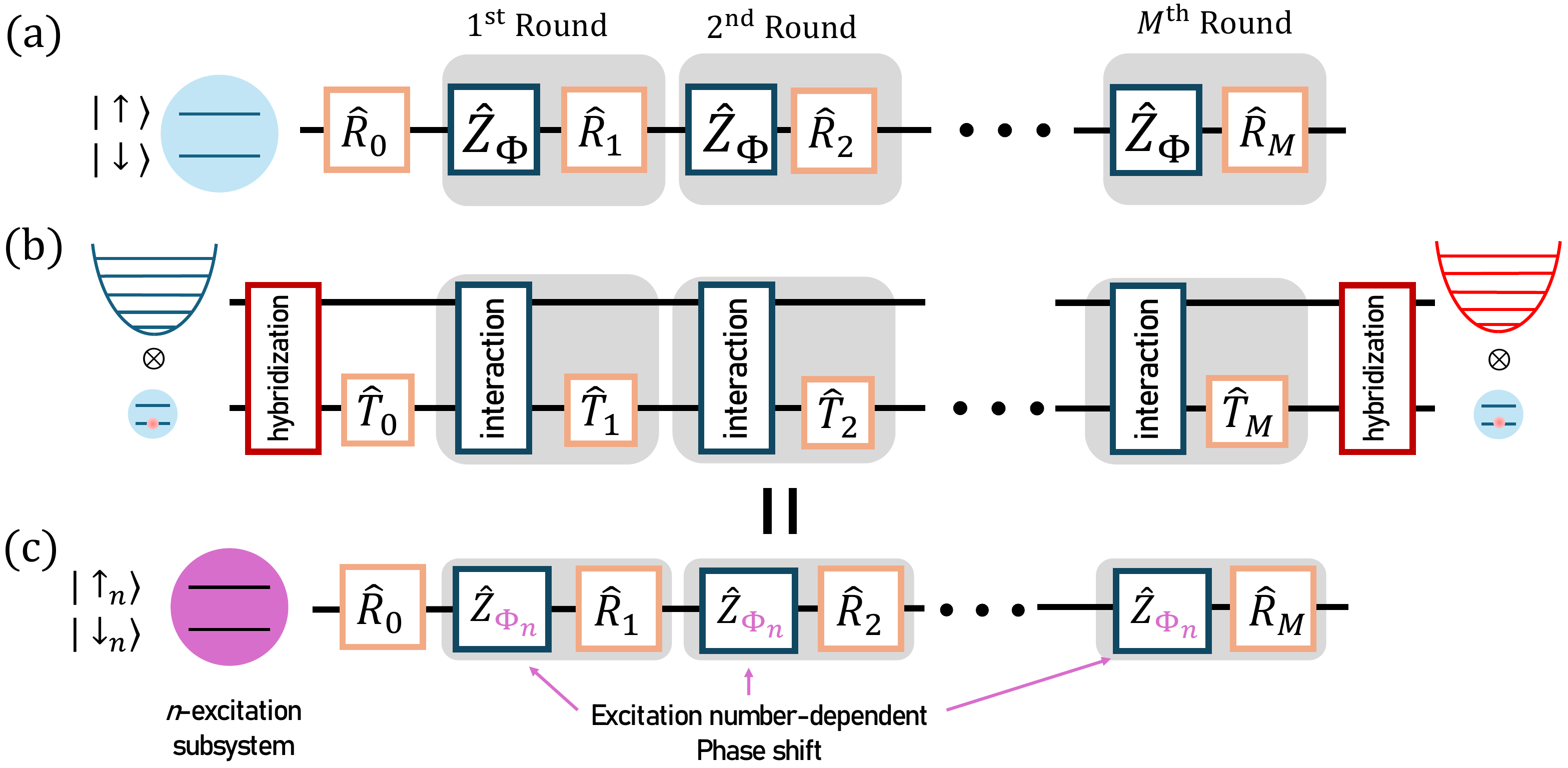}
    \caption{Illustration of QSP circuits. (a) QSP is a sequence of operations that alternatively applies fixed phase-shifts $\hat{Z}_\Phi$ and qubit rotations $\hat{R}_m$ for $M$ rounds.  (b) To engineer a qumode gate, qubit driving $\hat{T}_m$ and hybrid qumode-qubit interactions are applied alternatively for $M$ rounds. (c) The hybrid interaction induces excitation-number-dependent phase-shifts, such that each dressed subspace effectively undergoes QSP. After the sequence, the qubit returns to its ground state and is disentangled from the qumode.}
    \label{QSP_circuit}
\end{figure*}

\subsection{Gate Engineering through QSP}\label{sec_gate_QSP}

For many hybrid couplings, the joint qumode-qubit eigenstates can be recognized as a collection of two-level subspaces, each spanned by the dressed states $\{\ket{\downarrow_{n}}, \ket{\uparrow_{n}}\}$. Our strategy is to apply controls that drive transition only within each subspace. This allows us to use QSP to engineer the desired transformation on all dressed states. 

Fig.~\ref{QSP_circuit}(b) depicts the general scheme for QSP gate engineering. Consider the qumode is associated to a qubit with computational basis $\{\ket{g},\ket{e}\}$. The process begins with the qubit initialized in $\ket{g}$ and an arbitrary initial qumode state, $\ket{\psi}\equiv \sum_{n=0}^\infty c_n \ket{n}$, where $\ket{n}$ is the boson-number state with $n$ bosons and $c_n$ is the associated amplitude. In the first step, a hybridization process is performed to transform the computational basis states into the dressed states: 
\begin{eqnarray}\label{hybridization_eq}
\ket{n}\ket{g} \to  \ket{\downarrow_{n}} \quad \text{and} \quad \ket{n}\ket{e} \to  \ket{\uparrow_{n}}.
\end{eqnarray}

Next, a specific sequence of qubit driving $\hat{T}_m$ and hybrid interaction is employed. Particularly, the qubit drives are carefully selected to generate the same rotations $\hat{R}_m$ within every dressed subspace, but avoid inducing transitions between different subspaces.
On the other hand, the hybrid interaction provides the phase-shifts $\hat{Z}_{\Phi_n}$ that the shifted phase $z_n \equiv e^{i\Phi_{n}}$ depends on the boson number, which is distinct for each subspace. As a result, the sequence of qubit driving and hybrid interaction will generate a QSP sequence in every dressed subspace, as illustrated in Fig.~\ref{QSP_circuit}(c). Each dressed-state transforms distinctly as
\begin{eqnarray}\label{hybridization_general}
  \ket{\downarrow_{n}} &\xrightarrow[]{\text{oQSP}}& \Big[ F(z_n)\ket{\downarrow_{n}} + G(z_n)\ket{\uparrow_{n}}\Big], \quad \text{or}\nonumber\\
  \ket{\downarrow_{n}} &\xrightarrow[]{\text{GQSP}}&\Big[ P(z_n)\ket{\downarrow_{n}} + Q(z_n)\ket{\uparrow_{n}}\Big].
\end{eqnarray}

Finally, the hybridization in Eq.~\eqref{hybridization_eq} is reversed to restore the system to the computational basis states. Overall, the joint state of the hybrid system will be transformed as
\begin{eqnarray}
   \ket{n}\ket{g} &\xrightarrow[]{\text{oQSP}}&  F(z_n)\ket{n}\ket{g} + G(z_n)\ket{n}\ket{e}, \quad \text{or}\nonumber\\
  \ket{n}\ket{g} &\xrightarrow[]{\text{GQSP}}& P(z_n)\ket{n}\ket{g} + Q(z_n)\ket{n}\ket{e}. \label{general_result}
\end{eqnarray}

In general, transformation~\eqref{general_result} cannot be used as a bosonic gate because the qumode and qubit will become entangled.
Our objective is to design the polynomials to implement a non-linear phase gate in the form of $\sum_{n=0}^\infty e^{i\Theta_n}\ket{n}\bra{n}$, where the phase angles $\Theta_n$ of every boson-number state can be arbitrarily chosen. This can be accomplished by the polynomials that satisfy
\begin{eqnarray}
    F(z_n)&=&e^{i\Theta_n} \quad\text{(for oQSP)},\quad \text{or} \nonumber\\P(z_n)&=&e^{i\Theta_n} \quad \text{(for GQSP)}.
\end{eqnarray}
Then, the constraints \eqref{FG_constraint} and \eqref{PQ_constraint} inherently enforce the complementary polynomial to satisfy $G(z_n)=0$ and $Q(z_n)=0$. This ensures that the qubit returns to $\ket{g}$ and is hence disentangled with the qumode after the operation.

\section{Dispersive coupling systems}\label{sec_mod_k}

The first type of hybrid qumode-qubit system we consider is that with dispersive coupling, which induces a qubit frequency shift that is linearly dependent on the boson number. The corresponding Hamiltonian is given by \cite{Schuster2007}:
\begin{eqnarray} \label{hamilton_dis}
    \hat{H}_{\text{dis}} &=&  -\chi \hat{a}^\dag \hat{a} \ket{g}\bra{g},
\end{eqnarray}
where $\chi$ denotes the dispersive coupling strength. 
This type of system is commonly found in superconducting circuit QED \cite{Wallraff2004, Schuster2007}, as well as in other platforms, such as Rydberg atoms \cite{PhysRevA.45.5193} and trapped ions \cite{PhysRevA.55.2387}. 

The energy eigenstates of dispersive coupling are tensor products of the boson-number states and the qubit states. They form a collection of subspaces, each spanned by $\{\ket{\downarrow_n}=\ket{n}\ket{g},\ket{\uparrow_n}=\ket{n}\ket{e}\}$. Since the computational basis and the dressed basis coincide, no hybridization is required. 

To execute QSP, we utilize the coupling~\eqref{hamilton_dis} to generate the controlled-phase-shift,
\begin{eqnarray}
    \hat{C}_{\chi T} \equiv \exp{\left(i \chi T \hat{a}^\dag\hat{a}\ket{g}\bra{g}\right)}, \label{C_beta}
\end{eqnarray}
after an interaction duration $T$. It induces the phase-shift $\hat{Z}_{\Phi_n}$ to the dressed states, i.e. $\ket{\downarrow_n} \to e^{i\Phi_n}\ket{\downarrow_n}$ and $\ket{\uparrow_n}\to\ket{\uparrow_n}$, where the phase $\Phi_n= \chi T n$ linearly scales as the boson number $n$. 

Moreover, to generate the dressed-state rotation $\hat{R}_m$, Eq.~\eqref{rotation_matrix} up to a global phase, we consider the auxiliary qubit driving
\begin{eqnarray}
    \hat{T}_m = e^{i\left(\frac{\phi_m}{2}+\frac{\pi}{4}\right)\hat{\sigma}_z}e^{i \theta_m \hat{\sigma}_x}e^{i \left(\frac{\lambda_m}{2} + \frac{\pi}{4}\right) \hat{\sigma}_z},
\end{eqnarray}
where $\hat{\sigma}_x \equiv \ket{e}\bra{g}+\ket{g}\bra{e}$ and $\hat{\sigma}_z\equiv \ket{e}\bra{e}-\ket{g}\bra{g}$. Since every dressed subspace is a tensor product of qubit states and a fixed boson-number state, the qubit drive $\hat{T}_m$ preserves each dressed subspace and results in a homogeneous transformation in all subspaces. As dressed-state rotations are allowed along arbitrary axes, we can implement GQSP.

\subsection{Exponential-Parity Gate and QSP}\label{sec_ep}

Before introducing the new gates enabled by QSP, we first revisit a non-Gaussian gate that has been realized with dispersive coupling: the eParity gate \cite{PhysRevA.95.022303, ep_gate_2017, Gao2019}. It applies opposite phase-shifts to boson-number states based on their parity:
\begin{eqnarray}\label{U_ep}
    \hat{U}_{\text{eP}} 
    &\equiv&\sum_{\text{even}\,n} e^{i\Theta} \ket{n}\bra{n} + \sum_{\text{odd}\, n} e^{-i\Theta} \ket{n}\bra{n},
\end{eqnarray}
where $\Theta$ is an arbitrarily chosen phase angle.
Together with beam-splitters, this gate can form the exponential-SWAP gate, which allows universal DVQC with any bosonic encoding \cite{ep_gate_2017, Gao2019}. 

In Ref.~\cite{Gao2019}, the eParity gate is engineered by the following sequence of unitaries: $\hat{\mathcal{H}} \hat{C}_\pi \hat{X}_\Theta \hat{C}_\pi \hat{\mathcal{H}}$, where $\hat{\mathcal{H}}$ is the qubit Hadamard gate and $\hat{X}_\Theta = \exp\left( i \hat{\sigma}_x \Theta \right)$ is the qubit $x$-axis rotation. We recognize that the Hadamard gates and the $x$-rotation are also dressed-state rotations, and the controlled-phase-shift $\hat{C}_\pi$ induces a dressed-state phase shift $e^{i n \pi}$. Therefore, the gate engineering sequence coincides with the QSP sequence. Explicitly, it generates the transformation in Eq.~\eqref{general_result}
with $z_n = e^{i\pi n}$ and the polynomials
\begin{eqnarray}
    P(z) &=& \frac{\cos\Theta}{2} + i \sin\Theta z + \frac{\cos\Theta}{2}z^2,\\
    Q(z) &=& \frac{\cos\Theta}{2} - \frac{\cos\Theta}{2}z^2.
\end{eqnarray}

Since every boson-number state with the same parity constitutes the same value in the polynomial, i.e. $z_n = e^{i0}=1$ for even $n$ and $z_n=e^{i \pi}=-1$ for odd $n$, the gate will introduce a transformation that depends solely on the boson-number parity. Explicitly, we have $P(1) = e^{i\Theta}$ for even-parity and $P(-1) = e^{-i\Theta}$ for odd-parity. Because $Q(1)=Q(-1)=0$, the qubit becomes disentangled from the qumode afterwards and the eParity gate~\eqref{U_ep} is realized. 

\subsection{mod-$k$ Gate}

We recognize that the parity of an integer $n$ is determined by the value $n \,(\text{mod 2})$, i.e. $0$ for even and $1$ for odd. Therefore, a natural generalization of the eParity gate is a non-linear phase gate whose phase shift depends on the generalized-parity, $n \,(\text{mod} \,k)$ of the boson-number state $\ket{n}$. We call this new non-Gaussian gate the mod-$k$ gate, whose operator form is
\begin{eqnarray}
    \hat{U}_{\text{m}k} &=& \sum_{n=0}^\infty e^{i\Theta_{n \, (\text{mod} \, k)}} \ket{n}\bra{n},\label{U_modk}
\end{eqnarray}
where the phases $\Theta_{n\,(\text{mod}\, k)}$ can be arbitrarily chosen; $k$ can be any positive integer of choice.

To realize the mod-$k$ gate, we first choose the controlled-phase interaction duration as $T = 2\pi/(k\chi)$. This ensures that all boson-number states with the same generalized-parity contribute to the same phase shift in the QSP sequence, i.e. $z_n = z_{n \, (\text{mod} \, k)} =\omega_k^n$, where $\omega_k \equiv \exp(i2\pi/k)$.
Our task is to design a polynomial $P(z)$ that satisfies the condition~\eqref{constraint_of_P} and
\begin{eqnarray}\label{constraint}
    P(\omega_k^n) = e^{i\Theta_{n\, (\text{mod}\, k)}}.
\end{eqnarray} 

Before we present a systematic method to construct the desired polynomials, we note that our QSP scheme is different from most applications of QSP, which aim to design polynomials to approximate desired continuous functions (see examples in Ref.~\cite{QSP2021}). For our purpose of gate engineering, we want the polynomial to take exact values as required by Eq.~\eqref{constraint}, but only at a finite number of points $z_n$.

\subsection{Kernel Method}\label{sec_processing}

We propose to construct the polynomial as a sum of $k$ kernels:
\begin{eqnarray}\label{interpolation}
    P(z) = \sum_{n=0}^{k-1} e^{i\Theta_{n\, (\text{mod}\, k)}} K_n(z).
\end{eqnarray}
Here each kernel $K_n(z)$ is a polynomial that is unity at $z=\omega_k^n$ and zero at $z=\omega_k^m$ when $m\neq n$, i.e. $K_n(\omega_k^m) = \delta_{nm}$, where $\delta_{nm}$ is the Kronecker delta. Since in the case of dispersive coupling $z_n$'s are uniformly distributed on the unit circle, we can construct all kernels by ``rotating" that for $n=0$ on the phase space, i.e. 
\begin{eqnarray}\label{kn_k0}
    K_n(z) = K_0 (z \omega_k^{-n}).
\end{eqnarray}

In general, not all functions satisfying these conditions can be used as the kernel, because Eq.~\eqref{constraint_of_P} may not hold for all $z$ on the unit circle. Here, we introduce a kernel that satisfies all the desired properties and has a degree $M=2k$ only:
\begin{eqnarray}\label{kernel_form}
    K_0(z) = \frac{(z+1)^2}{4 } \frac{\prod^{k-1}_{m=1}(z-\omega_k^m)^2}{\prod^{k-1}_{m=1}(1-\omega_k^m)^2}.
\end{eqnarray}
In Appendix~\ref{appendix_bound}, we prove that any polynomial $P(z)$ constructed by our method must satisfy condition~\eqref{constraint_of_P}. Once $P(z)$ is determined, the corresponding polynomial $Q(z)$ and the associated rotation angles can be derived, for example, by using the algorithm proposed in Ref.~\cite{GQSP2024}.

We now discuss the implementation time of the mod-$k$ gate. Since the required polynomial has a degree of $2k$, only $2k$ rounds of the controlled-phase-shift and rotation are necessary. In the systems where the dispersive coupling is much weaker than the single-qubit drive, the operation time is dominated by that of the controlled-phase-shifts. Since each round takes a time of $T=2\pi/(k\chi)$, the total implementation time is given by
\begin{eqnarray}\label{total_time}
    T_{\text{total}} = 4\pi/\chi.
\end{eqnarray}
Surprisingly, it is independent of both $k$, the number of independent phases to be induced, and the number of boson levels involved. Since our construction is exact, our formulation of the mod-$k$ gates does not suffer from the precision-duration trade-off.
 
\section{Applications of mod-$k$ gates}\label{sec_app}

\subsection{SNAP Gate}\label{sec_app_SNAP}

A commonly employed non-Gaussian gate in circuit QED platforms is the SNAP gate \cite{SNAP_2015}, which applies distinct phases to every boson-number states based on their boson number. Its operator form is given by
\begin{eqnarray}\label{equation_snap}
    \hat{U}_{\text{SNAP}} &=& \sum_{n=0}^{N_{\text{max}}} e^{i\Theta_{n}} \ket{n}\bra{n} + \text{other},\label{U_SNAP}
\end{eqnarray}
where ``other" denotes the terms associated with the states with higher boson number than the maximum of concern, $N_{\text{max}}$.  

In the conventional implementation of SNAP gates, the qubit drive incorporates multiple frequency tones, each resonating with the energy gap between a particular dressed-state pair $\{\ket{n}\ket{g},\ket{n}\ket{e}\}$ (see Appendix~\ref{appendix_multi} for details). This method allows each boson-number state to be independently manipulated. However, to avoid unwanted off-resonant effects, the amplitude of each tone must remain much weaker than that of the dispersive coupling. This results in a prolonged implementation time $T_{\text{total}} \gg 1/\chi$.

\begin{figure}
    \centering
    \includegraphics[scale=0.32]{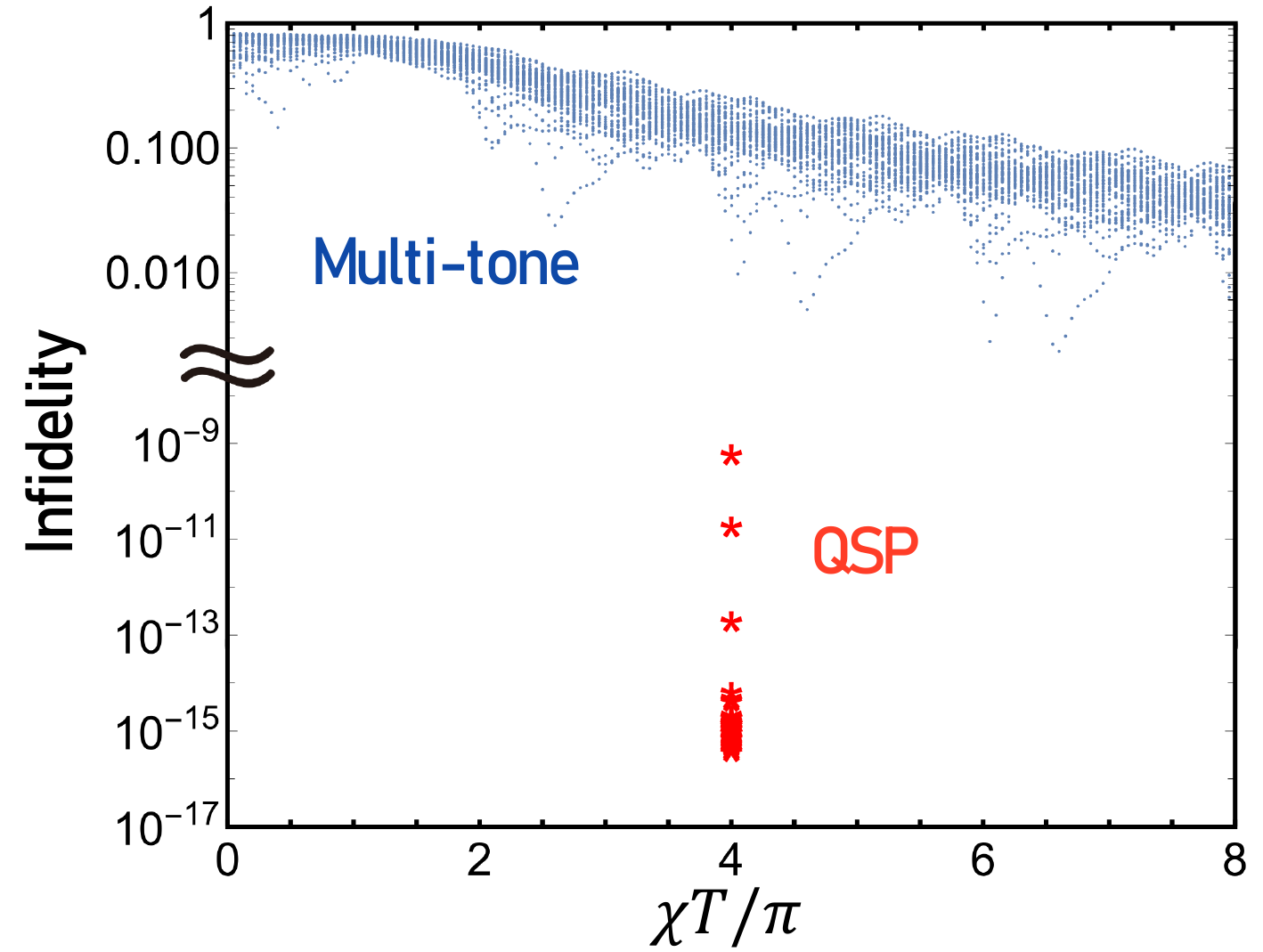}
    \caption{Gate infidelities of both the multi-tone driving (blue dots) and QSP (red dots) implementations of SNAP gates with $N_{\text{max}}=4$. The details of the fidelities and simulations can be found in Appendix~\ref{appendix_gate_fidelity} and \ref{appendix_multi}, respectively. The presented data consists of $50$ samples of SNAP gates with randomly chosen target phases. For the multi-tone driving method, the infidelities decrease gradually with implementation time, because a weaker drive implements the gate more slowly but reduces the off-resonant effects. In contrast, the QSP method implements the same gates in a fixed time of $T_{\text{total}} = 4\pi/\chi$, but their infidelities are orders of magnitude lower.}
    \label{fig1}
\end{figure}

The mod-$k$ gate can be reduced to the SNAP gate by setting $k = N_{\text{max}} +1$ and engineering $\Theta_{n\,\text{mod}\, k} =\Theta_{n}$. Compared to the conventional method, implementing SNAP gates with our QSP method can be both higher in accuracy and faster. To demonstrate the merits of our method, we simulate the implementation of SNAP gates for $N_{\text{max}}=4$, five bosonic levels. As shown in Fig.~\ref{fig1}, the conventional method typically exhibits infidelity greater than $10^{-2}$ for the implementation time up to $T_{\text{total}} = 8\pi/\chi$. In contrast, our QSP method engineers the same gate in a fixed time of $4\pi/\chi$, while the infidelity is at the order of $10^{-9}$ in the worst case we have tested. A detailed example of a SNAP gate that modifies only the phase of the $0$-Fock state $\ket{0}$ is demonstrated in Appendix.~\ref{appendix_gate_example}. We note that the inaccuracy originates primarily from numerical errors in our angle-finding algorithm, which can be further reduced with improved computational machinery.

\subsection{Generic Qudit Entangling Gate}\label{sec_app_qudit}

Due to the enlarged Hilbert space, using qudits for QIP could reduce circuit complexity and simplify experimental setups \cite{Chi2022, RevModPhys.97.021003}. However, conventional implementations of qudit entangling gates often concatenate multiple two-level operations \cite{Ringbauer2022, Hrmo2023}, so the total implementation time increases with the qudit dimension.
Here, we propose to use QSP to engineer a generic qudit entangling gate, whose implementation time is independent of the qudit dimension.
To achieve this, we first generalize the mod-$k$ gate to a two-mode scenario, which is defined as
\begin{eqnarray}\label{Tmod}
    \hat{U}_{\text{Tm}k} 
    =\sum_{n_A=0}^\infty\sum_{n_B=0}^\infty &&\exp\left(i\Theta_{n_T \,(\text{mod}\,k)}\right) \nonumber\\
    &&\times \ket{n_A,n_B}\bra{n_A,n_B},
\end{eqnarray}
where $n_A$ and $n_B$ denote the boson number of the qumodes A and B respectively, and $n_T \equiv n_A + n_B$ is the total number of bosons in the two qumodes. Then, we will demonstrate that for boson-number state encoding, this gate is a qudit CPhase gate \cite{qudit_cluster_CP_2003}, which is the building block for qudit cluster states \cite{qudit_cluster_CP_2003}. 

This gate can be implemented by coupling both qumodes to the same qubit through a dispersive coupling with the same strength, i.e.
\begin{eqnarray}\label{twomodeH}
    \hat{H}_I = - \chi \left( \hat{a}^\dag \hat{a} +  \hat{b}^\dag \hat{b}\right) \ket{g}\bra{g},
\end{eqnarray} 
where $\hat{a}$ and $\hat{b}$ are the annihilation operators of the qumodes A and B respectively. Evolving under this Hamiltonian is equivalent to applying controlled-phase-shifts onto both qumodes; the evolution can be expressed as $\hat{C}^{A}_{\chi T}\hat{C}^{B}_{\chi T}$. This generates the phase-shift $\hat{Z}_{\Phi_n}$ in the two-mode dressed basis $\{\ket{\downarrow_{n_A,n_B}}=\ket{n_A}\ket{n_B}\ket{g},\ket{\uparrow_{n_A,n_B}}=\ket{n_A}\ket{n_B}\ket{e}\}$, where $\Phi_{n} = \chi T n_T$ scales linearly as the total boson number $n_T$.

\begin{figure}
    \centering
\includegraphics[scale=0.23]{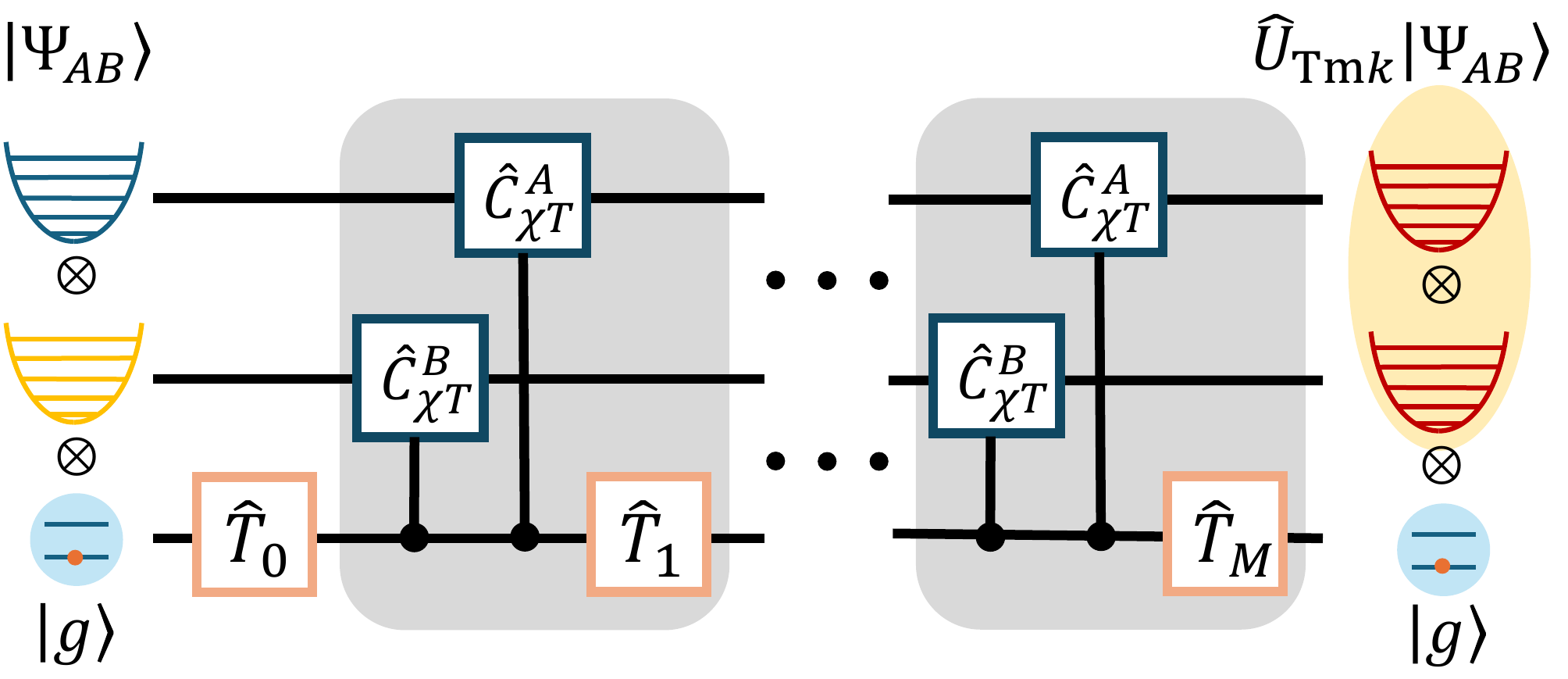}
    \caption{Quantum circuit of the two-mode mod-$k$ gate. Both qumodes A and B are dispersively coupled to the qubit and undergo controlled-phase-shifts. 
   By applying well-designed qubit rotations between each round of controlled-phase-shifts, the two-mode mod-$k$ gate is applied to the qumodes.}
    \label{circuit2}
\end{figure}

After applying a sequence of qubit drivings and controlled-phase-shifts with $\chi T = 2\pi/(2d-1)$, as shown in Fig.~\ref{circuit2}, each dressed state is transformed as
\begin{eqnarray}
    \ket{n_A}\ket{n_B}\ket{g} &\to& P((\omega_{2d-1})^{n_T})\ket{n_A}\ket{n_B}\ket{g} \nonumber\\
    &+& Q((\omega_{2d-1})^{n_T})\ket{n_A}\ket{n_B}\ket{e}.
\end{eqnarray}
The two-mode mod-$k$ gate is implemented by designing polynomials to satisfy $P((\omega_{2d-1})^{n_T})= \exp\left( i \Theta_{n_T \,(\text{mod}\,(2d-1))} \right)$, which also implies $Q((\omega_{2d-1})^{n_T})=0$. Since this condition is the same as the single-mode case, cf. Eq.~\eqref{constraint}, the kernel method described in Sec.~\ref{sec_processing} can be directly applied with the substitutions $n \to n_T$ and $k \to (2d-1)$. Since the degree of the polynomial is $2(2d-1)$ and each round of dispersive interaction requires a duration of $2\pi/[(2d-1)\chi]$, the total implementation time remains $4\pi/\chi$, which is independent of the qudit dimension.

For the qudit encoding that the computational bases are the boson-number states with $n=0,1,...,d-1$ bosons, the two-mode mod-$k$ gate is equivalent to a qudit CPhase gate \cite{qudit_cluster_CP_2003} when the phases are engineered as
\begin{eqnarray}
   \Theta_{n_T \,(\text{mod}\,(2d-1))} = \frac{n_T^2}{d}\pi ,
\end{eqnarray}
where $n_T =0,1,...,2d-2$. This choice of parameters can be understood by first expanding $n_T^2 = n_A^2+n_B^2+2n_A n_B$. Substituting this into Eq.~\eqref{Tmod} and replacing the photon numbers by the corresponding operators, $n_A \to \hat{a}^\dag\hat{a}$ and $n_b \to \hat{b}^\dag\hat{b}$, we see that this gate takes the form of a CPhase gate \cite{qudit_cluster_CP_2003}, $\exp \left( i2\pi \hat{a}^\dag\hat{a} \hat{b}^\dag\hat{b}/d \right)$, up to some local non-linear phase-shifts $\exp \left( i\pi \left[(\hat{a}^\dag\hat{a})^2 +(\hat{b}^\dag\hat{b})^2\right] /d \right)$. 

\subsection{Logic Gate for Rotation-Symmetric Codes}\label{sec_app_rotation}

Another application of the mod-$k$ gate is as a logic gate for rotation-symmetric bosonic codes \cite{rotation_symmetric_2020}. These codes can protect quantum information against loss, gain and dephasing noises \cite{rotation_symmetric_2020}. Some codes belonging to this family, such as cat \cite{cat1999} and binomial codes \cite{bosonic2016}, have recently been demonstrated to achieve the break-even point \cite{Ofek2016, Hu2019, doi:10.1126/science.aat3996}. 

These codes leverage the states with discrete rotational symmetry. An $L$-fold rotation-symmetric state is a superposition of the boson-number states in which the excitation numbers are multiples of an integer $L$. The logical basis states of an $L$-fold rotation-symmetric qubit are generally given by 
\begin{eqnarray}
    \ket{0_L} &=& \sum_{l=0}^\infty f_{2lL} \ket{2lL}, \quad \text{and}\\
    \ket{1_L} &=& \sum_{l=0}^\infty f_{(2l+1)L} \ket{(2l+1)L},
\end{eqnarray}
where the superposition amplitudes $f_n$ satisfy the normalization condition. The code distance can be increased by using a larger $L$. 

Now we discuss how to use mod-$k$ gates to implement the logic gates for these codes. 
First, we consider the single-qubit phase gate, defined as:
\begin{eqnarray}
    \ket{0_L} \to \ket{0_L} \quad \text{and} \quad \ket{1_L} \to e^{i\Theta}\ket{1_L}.
\end{eqnarray}
This requires that the boson-number states are transformed as $\ket{2lL} \to \ket{2lL}$ and $\ket{(2l+1)L} \to e^{i\Theta}\ket{(2l+1)L}$ with $l=0,1,2,...$. We can use a mod-$k$ gate with $k=2L$ and a polynomial $P(z)$ that satisfies $P(\omega_{2L}^n) = 1$ for $n = 2lL $ and $P(\omega_{2L}^n) = e^{i\Theta}$ for $ n= (2l+1)L$.

One might also want the gate to apply the same logical transformation even if the state has suffered from $L_{\text{loss}}$ excitations loss or $L_{\text{gain}}$ excitations gain, where $L_{\text{loss}}, L_{\text{gain}}< \lfloor L/2 \rfloor$. This can be achieved by further requiring the polynomial to satisfy 
\begin{eqnarray}\label{poly_error}
    P(\omega_{2L}^n) = 
    \begin{cases}
    1 & \text{for} \quad  n = 2lL -\lfloor \frac{L}{2}\rfloor,..., \\
    & \quad \quad \quad \quad  2lL+ \lfloor  \frac{L}{2}\rfloor \\
    e^{i\Theta} & \text{for} \quad n = (2l+1)L -\lfloor \frac{L}{2}\rfloor,..., \\
    & \quad \quad \quad \quad   (2l+1)L+ \lfloor  \frac{L}{2}\rfloor  
    \end{cases}.\nonumber\\
\end{eqnarray}

Similarly, the logic CPhase gate can be realized by a two-mode mod-$k$ gate with $k=2L$, which can be engineered via QSP with the polynomial $P(z)$ satisfying Eq.~\eqref{poly_error}. This gate applies the following phase shifts on the logical basis states :
\begin{eqnarray}
   && \ket{0_L}\ket{0_L} \to \ket{0_L}\ket{0_L} \, , \,
   \ket{1_L}\ket{0_L} \to e^{i\Theta}\ket{1_L}\ket{0_L},  \nonumber\\
    &&\ket{0_L}\ket{1_L} \to \ket{0_L} \ket{\tilde{1}_L} \, , \,\ket{1_L}\ket{1_L} \to e^{-i\Theta}\ket{1_L}\ket{\tilde{1}_L},\nonumber\\
\end{eqnarray}
where $\ket{\tilde{1}_L} = e^{i\Theta}\ket{1_L}$. Notably, our method can be extended straightforwardly to engineer logic gates for qudit codes.

\subsection{Deterministic Multiple-Component Cat State Generation} \label{sec_app_cat}

In addition to DV QIP, the mod-$k$ gate can also be used in CV QIP. A particular application is to coherently generate multiple-component cat states \cite{PhysRevA.45.6570}. These states have been proposed as resources for observing the quantum-to-classical transition \cite{PhysRevLett.77.4887}, testing the foundations of quantum theory \cite{PhysRevA.102.022202, PhysRevA.45.6811, PhysRevA.67.012105,PhysRevA.67.012106}, and quantum metrology \cite{Zurek2001, PhysRevA.73.023803}.

Conventional approaches to prepare cat states usually require measurements; this makes the schemes non-deterministic. 
For example, in photonic platforms, a cat state can be probabilistically generated by first splitting a non-vacuum state by a beam splitter, and then performing homodyne or photon number measurement in one mode \cite{PhysRevA.55.3184, Ourjoumtsev2007}. Furthermore, in cavity QED platforms, the qumode is first entangled with the qubit through conditional displacement gates \cite{Hacker2019, PhysRevA.105.063717}. Then, by measuring the qubit, the qumode will collapse to either an even- or odd-symmetric cat state. 
In contrast, our approach requires only coherent interaction and is thus deterministic. Besides, the cat states generated by the above conventional methods have only two components, while our method can generate a cat state with an arbitrary number of components.

We recognize that the mod-$k$ gate can always be expressed as a superposition of $k$ bosonic rotation, $\hat{U}_R(\nu) \equiv e^{i\nu\hat{a}^\dag\hat{a}}$, and the rotation angles are given by $\nu=2\pi l/k $ $(l=0,1,2,..,k-1)$, i.e.
\begin{eqnarray}\label{superposition_of_rotation}
    \hat{U}_{\text{m}k} = \sum_{l=0}^{k-1} \xi_l \hat{U}_R \left( \frac{2\pi}{k}l \right).
\end{eqnarray}
Here, the coefficients $\xi_l = p_l+p_{l+k}+ \delta_{0l}p_{l+2k}$ are sums of the polynomial coefficients $p_l$ (cf. Eq.~\eqref{eq_of_p_q}). The intuition is that the mod-$k$ gate is equal to a polynomial $P(\hat{z})$ of the operator $\hat{z}\equiv e^{i2\pi \hat{a}^\dag\hat{a}/k}$, which is a rotation with angle $2\pi/k$. 

When the mod-$k$ gate is applied to a coherent state $\ket{\alpha}$,  it generates a $k$-component cat state $\sum_{l=0}^{k-1} \xi_l\ket{\alpha e^{i 2\pi l/k }}$. 
Particularly, we find that the equal-amplitude $k$-component cat state,
\begin{eqnarray}\label{cat_eq_eq}
\ket{\text{CAT}_{k}} = \sum_{l=0}^{k-1} \frac{e^{i\frac{l(l-k) \pi}{k}}}{\sqrt{k}}  \ket{\alpha e^{i\frac{2\pi}{k}l}},
\end{eqnarray}
can be generated by choosing the target phases:
\begin{eqnarray}\label{relation}
    e^{i\Theta_{n \,\text{mod} \, k}} = \sum_{l=0}^{k-1} e^{i\frac{l(l-k) \pi}{k}}e^{i\frac{2\pi}{k} l n}/\sqrt{k}.
\end{eqnarray}
Here, the target phases are determined by substituting $\xi_l=e^{i\frac{l(l-k) \pi}{k}}/\sqrt{k}$ into Eq.~\eqref{superposition_of_rotation}. An example of a $5$-component cat state generated by a mod-$5$ gate is illustrated in Fig.~\ref{fig_cat}, and the corresponding QSP polynomials and rotation angles are listed in Appendix~\ref{appendix_gate_example}.

\begin{figure}
    \centering
    \includegraphics[scale=0.23]{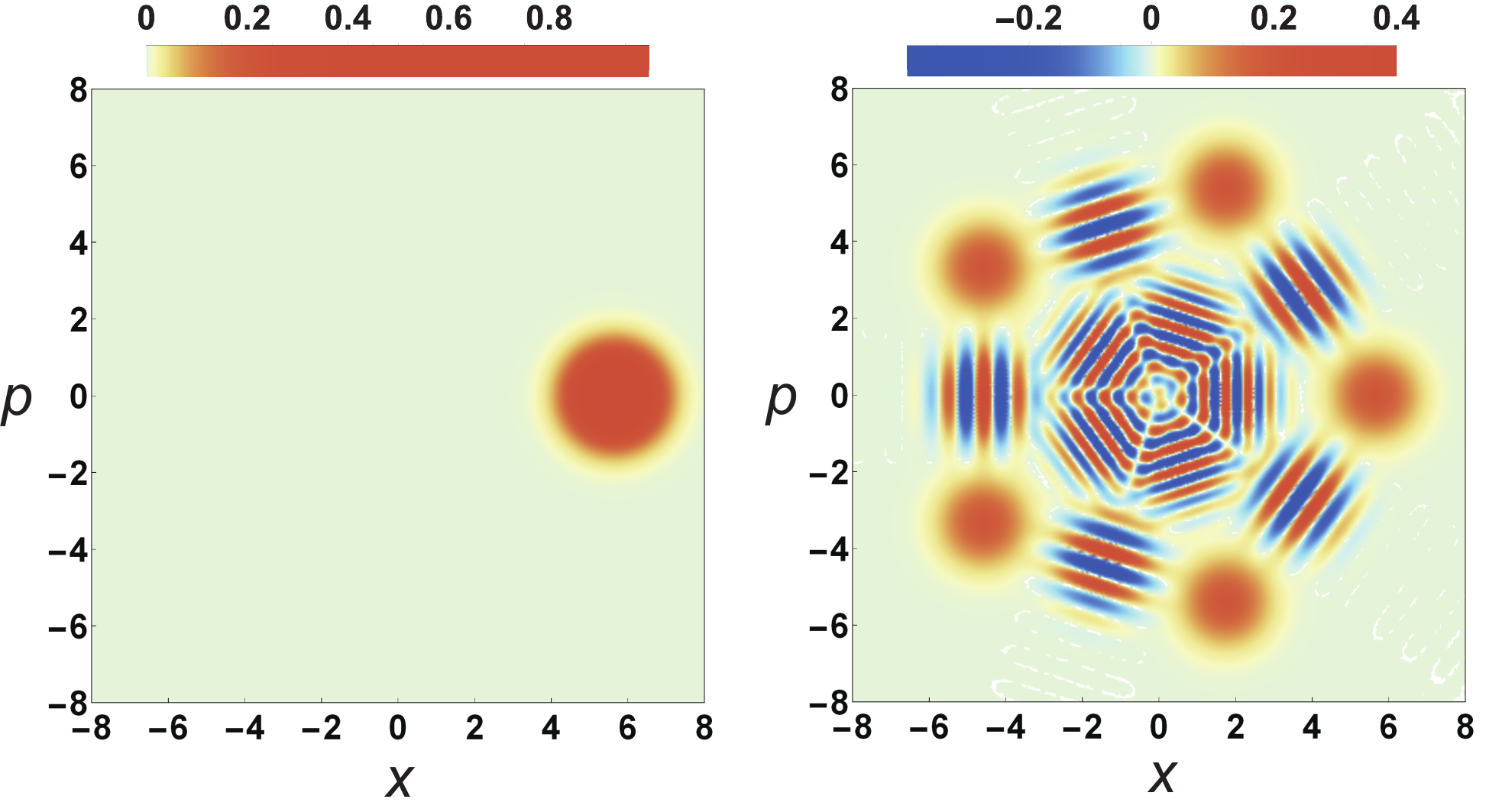}
    \caption{Wigner functions of (left) a coherent state with amplitude $\alpha =4$ and (right) an equal-amplitude $5$-component cat state generated by using the mod-$5$ gate with the phases in Eq.~\eqref{relation}.}
    \label{fig_cat}
\end{figure}

\section{Jaynes-Cummings Systems}\label{sec_JC}

Apart from dispersive coupling, JC interaction is another common type of hybrid interaction \cite{JC_1963}. It can be found in many quantum platforms, such as trapped ions \cite{RevModPhys.75.281} and cavity QED systems \cite{RevModPhys.87.1379}. The JC interaction creates a qubit excitation by annihilating a bosonic excitation and the other way round. Its Hamiltonian is given by
\begin{eqnarray}\label{Ham_JC}
    \hat{H}_{\text{JC}} = \lambda \left(\hat{a} \ket{e}\bra{g} + \hat{a}^\dag \ket{g}\bra{e}\right),
\end{eqnarray}
where $\lambda$ is the JC interaction strength. Diagonalizing the JC Hamiltonian, we obtain the corresponding dressed-state pairs:
\begin{eqnarray}\label{dressed_state_JC}
    \ket{\downarrow_{n}} &=& \frac{1}{\sqrt{2}} \left(\ket{n+1}\ket{g} - \ket{n}\ket{e}\right) \quad \text{and} \nonumber\\
    \ket{\uparrow_{n}} &=&  \frac{1}{\sqrt{2}} \left(\ket{n+1}\ket{g} + \ket{n}\ket{e}\right),
\end{eqnarray}
for integer $n \in[0,\infty)$. We note that the ground state $\ket{0}\ket{g}$ will not form any dressed pair. 

To engineer gates via QSP, we first transform the computational basis states to the dressed states via hybridization; this will be discussed in Sec.~\ref{hybridization}. Here, we first assume that hybridization has been accomplished and focus on how phase-shifts and rotations are implemented and how the required polynomials are constructed.

Evolving under the JC Hamiltonian, each dressed state will acquire a phase shift that depends non-linearly on the boson number. In terms of the dressed basis $\{\ket{\downarrow_{n}} \equiv (1,0)^T, \ket{\uparrow_{n}}\equiv (0, 1)^T\}$, this evolution is expressed as
\begin{eqnarray}\label{JC_phase}
\begin{pmatrix}
e^{i\Phi_n/2} & 0 \\
0 & e^{-i\Phi_n/2}
\end{pmatrix} =
e^{-i\Phi_n/2} \hat{Z}_{\Phi_n}, 
\end{eqnarray}
where $\Phi_n = \lambda T \sqrt{n+1}$.
This comprises a phase-shift operator $\hat{Z}_{\Phi_n}$ that can be used in QSP gate engineering, and an additional phase $e^{-i\Phi_n/2}$. We note that this phase is not a global phase, as its boson-number dependence will introduce a relative phase between components with different boson numbers. Therefore, it has to be eliminated when we want to apply the evolution~\eqref{JC_phase} in QSP. We will discuss this in the next subsection. 

Unlike in dispersive coupling systems, a general qubit rotation can cause transitions between different dressed subspaces in a JC system. We find that only the qubit $z$-axis rotation can retain the dressed states within the same subspace. This operation can be implemented by introducing a detuning $\Delta$ between the qumode and the qubit; the Hamiltonian is given by $\Delta \hat{\sigma}_z$ in the interaction picture. Evolving under the detuning will Rabi oscillate the dressed states, i.e.
\begin{eqnarray}
 \ket{\downarrow_n}&\to&  \cos\Delta t \ket{\downarrow_n}+ i\sin\Delta t\ket{\uparrow_n} \quad \text{and} \nonumber  \\
    \ket{\uparrow_n} &\to& i\sin\Delta t\ket{\downarrow_n} + \cos\Delta t \ket{\uparrow_n} ,
\end{eqnarray}
where $t$ is the evolution time.
This is effectively an $x$-axis rotation in the dressed-state basis. Since only single-axis rotations can be implemented, oQSP should be considered here (cf. Eq.~\eqref{QSP_sequence_1}). 
 
\subsection{Kernel Method for JC Systems}\label{sec_JC_kernel}

For JC systems, the transformation of $\ket{\downarrow_n}$ is given by Eq.~\eqref{hybridization_general} with the modification:
\begin{eqnarray}
    F(e^{i\Phi_n}) &\to& e^{-i\Upsilon_n}F(e^{i\Phi_n}) \quad \text{and}\\
    G(e^{i\Phi_n}) &\to& e^{-i\Upsilon_n}G(e^{i\Phi_n}),
\end{eqnarray}
where $\Upsilon_n \equiv M\lambda T\sqrt{n+1}/2$ is the additional phase in Eq.~\eqref{JC_phase} accumulated after $M$ rounds of JC interaction. If we aim to engineer the SNAP gate (cf. Eq.~\eqref{equation_snap}), the polynomials can be designed to eliminate the additional phases, 
\begin{eqnarray}
    F(e^{i\Phi_n})&=& e^{i\Upsilon_n}e^{i\Theta_n} \quad \text{and} \nonumber\\
    G(e^{i\Phi_n})&=& 0.
\end{eqnarray}

We again adopt the kernel method to design the desired polynomials. However, since the polynomial coefficients for oQSP must be real, we cannot use the kernel in Eq.~\eqref{kernel_form} that is designed for GQSP whose coefficients can be complex. Our approach here is to construct two kernels, one for the real part and the other for the imaginary part:
\begin{eqnarray}\label{kernel_modified}
     F(z) &=& \sum_{n=0}^{N_{\text{max}}} [\cos\Theta_n \mathcal{K}_n^R(z) + i\sin\Theta_n \mathcal{K}_n^I(z)].
\end{eqnarray}
These kernels $\mathcal{K}_n^R(z)$ and $\mathcal{K}_n^I(z)$ must be polynomials of $z$ with only real and imaginary coefficients, respectively. Moreover, the kernels are designed to satisfy $\mathcal{K}_n^{R(I)}(e^{i\Phi_m}) = e^{i\Upsilon_n}\delta_{nm}$. 
We observed that the polynomial degree of the kernels can be reduced by arranging $\Phi_n$ to be more evenly separated on the unit circle of the complex plane. Here, we set $\Phi_n = \pi \sqrt{n+1}/(2\sqrt{N_{\text{max}}+2})$ by choosing $T = \pi /(2\lambda\sqrt{N_{\text{max}}+2})$.

To satisfy the above requirements, we construct the kernels:
\begin{eqnarray}
    \mathcal{K}_n^R(z)& =&  e^{i\Upsilon_n}\frac{(e^{2i\delta_n^R}z^2+1)^h +(z^2+e^{2i\delta_n^R})^h }{(e^{2i\delta_n^R}e^{i2\Phi_n}+1)^h +(e^{i2\Phi_n}+e^{2i\delta_n^R})^h} \nonumber\\
   &\times& \prod_{m\neq n}^{N_{\text{max}}} \left[\frac{(z^2-e^{-i2\Phi_m}) (z^2-e^{i2\Phi_m})}{(e^{i2\Phi_n}-e^{-i2\Phi_m}) (e^{i2\Phi_n}-e^{i2\Phi_m})}\right]^s,\nonumber\\
   \mathcal{K}_n^I(z) &=& e^{i\Upsilon_n}\frac{(e^{2i\delta_n^I}z^2+1)^h -(z^2+e^{i\delta_n^I})^h }{(e^{2i\delta_n^I}e^{i2\Phi_n}+1)^h -(e^{i2\Phi_n}+e^{2i\delta_n^I})^h}\nonumber\\
   &\times& \prod_{m\neq n}^{N_{\text{max}}} \left[\frac{(z^2-e^{-i2\Phi_m}) (z^2-e^{i2\Phi_m})}{(e^{i2\Phi_n}-e^{-i2\Phi_m}) (e^{i2\Phi_n}-e^{i2\Phi_m})}\right]^s.\nonumber\\\label{kernel_modified_I}
\end{eqnarray}
Here, the degrees of both polynomials are $M = 4s N_{\text{max}}+2h$ and the parameters $\delta_n^{R(I)}$ are chosen to ensure $\mathcal{K}^{R(I)}_n(e^{i\Phi_n})=e^{i\Upsilon_n}$. In Appendix~\ref{appendix_JC}, we discuss how to determine the integer powers $h$ and $s$ by setting $\sum_{n=0}^{N_{\text{max}}} [|\mathcal{K}_n^R(z)|+|\mathcal{K}_n^I(z)|] \leq 1$ to ensure $|F(z)| \leq 1$. Generally, larger values of $h$ and $s$ are required for a larger $N_{\text{max}}$. 

\subsection{Hybridization}\label{hybridization}

Now, we discuss how to implement the hybridization that transforms the computational basis states to the dressed states, cf. Eq.~\eqref{hybridization_eq}.
Our scheme begins by initializing the qubit as $\ket{e}$, and the first hybridization aims to map $\ket{n}\ket{e}$ to $\ket{\downarrow_n}$. We recognize that the states $\ket{n}\ket{e}$ and $\ket{\downarrow_n}$ are respectively the eigenstates of the qumode-qubit detuning and JC interaction. Hybridization can, in principle, be achieved by multi-tone driving or adiabatically, i.e. JC interaction is always-on while the detuning is large at the beginning and then gradually reduced to zero. However, these processes are generally slow. Here, we propose an alternative approach to implement hybridization with QSP.

Since the computational basis state can be expressed as a superposition of the dressed states: $\ket{n}\ket{e} = (\ket{\downarrow_n}-\ket{\uparrow_n})/\sqrt{2}$, hybridization is effectively a Hadamard gate applied to each dressed subspace. This gate can be realized if we design the polynomials to satisfy $F(e^{i\Phi_n}) = G(e^{i\Phi_n})=e^{i\Upsilon_n}/\sqrt{2}$. While the condition for $F(z)$ can always be satisfied by using the real-part kernel,
\begin{eqnarray}
    F(z) = \sum_{n=0}^{N_\text{max}} \frac{1}{\sqrt{2}}\mathcal{K}^R_n(z),
\end{eqnarray}
a suitable $G(z)$ may not be easily designed because the existing algorithm can only determine the norm but not the phase of $G(z)$ \cite{chao2020findinganglesquantumsignal}. More explicitly, the existing algorithm can be used to design a $G(z)$ that satisfies $G(e^{i\Phi_n}) = e^{i\Upsilon_n}e^{i\zeta_n}/\sqrt{2}$ with a known but unwanted phase $e^{i\zeta_n}$. If we are using these polynomials, the dressed basis states are transformed as
\begin{eqnarray}\label{undesired_FG}
   \ket{\downarrow_n} &\to& \frac{1}{\sqrt{2}}\left(\ket{\downarrow_n} + e^{i\zeta_n}\ket{\uparrow_n} \right)\quad \text{and}\nonumber\\
   \ket{\uparrow_n} &\to& \frac{1}{\sqrt{2}}\left(-e^{-i\zeta_n}\ket{\downarrow_n} + \ket{\uparrow_n} \right).
\end{eqnarray}
As a consequence, the computational basis state $\ket{n}\ket{e} $ is transformed into a superposition of dressed states, $e^{-i\zeta_n/2} \cos(\zeta_n/2)\ket{\downarrow_n}+ie^{i\zeta_n/2}\sin(\zeta_n/2)\ket{\downarrow_n}$, instead of our desired $\ket{\downarrow_n}$.

To eliminate the effect of these unwanted phases, before the transformation~\eqref{undesired_FG}, we employ a gate in Sec.~\ref{sec_JC_kernel} to introduce the following phase shift to every dressed state: 
\begin{eqnarray}
    \ket{\downarrow_n} \to e^{i\zeta_n/2}\ket{\downarrow_n} \ \text{and} \  \ket{\uparrow_n} \to e^{-i\zeta_n/2}\ket{\uparrow_n}.
\end{eqnarray}
Overall, the hybridization~\eqref{hybridization_eq} is accomplished except that there will be some extra phases, $\ket{n}\ket{e} \to e^{i\zeta_n/2}\ket{\downarrow_n}$ and $\ket{n+1}\ket{g} \to e^{-i\zeta_n/2}\ket{\uparrow_n}$. However, these phases will not affect the phase gate engineering and can ultimately be cancelled by reversing the hybridization in the final step.

\subsection{Phase Gate Engineering}\label{appendix_example}

The gate engineering procedure comprises three steps, with a total of five rounds of QSP. The first step, which involves two rounds of QSP, is the hybridization that maps the computational basis to the dressed basis. The second step is to apply a single round of QSP to implement the desired dressed-state phase shift. The last step is to reverse the hybridization using two rounds of QSP, mapping the state back to the computational basis. The overall process can be written as
\begin{eqnarray}
    \ket{n}\ket{e} &\xrightarrow[]{\text{hybrid.}}& e^{i\zeta_n/2}\ket{\downarrow_n} \\
    &\xrightarrow[]{\text{Phase}}& e^{i\Theta_n}e^{i\zeta_n/2}\ket{\downarrow_n} \xrightarrow[]{\text{hybrid.}}  e^{i\Theta_n}\ket{n}\ket{e}.\nonumber
\end{eqnarray} 

The whole procedure takes five rounds of QSP, and each consists of $4sN_{\text{max}}+2h$ rounds of operations.
In the typical situation where the qubit drive is much stronger than the hybrid interaction, the operation time is dominated by that of the evolution under the JC interaction. We choose that each operation round takes $T = \pi/(2\lambda\sqrt{N_{\text{max}}+2})$, and the total implementation time is 
\begin{eqnarray}\label{JC_total_time}
    T_{\text{total}} 
    &=& \frac{5\pi(2sN_{\text{max}}+h)}{\lambda\sqrt{N_{\text{max}}+2}}.
\end{eqnarray}

\begin{figure}
    \centering
    \includegraphics[scale=0.3]{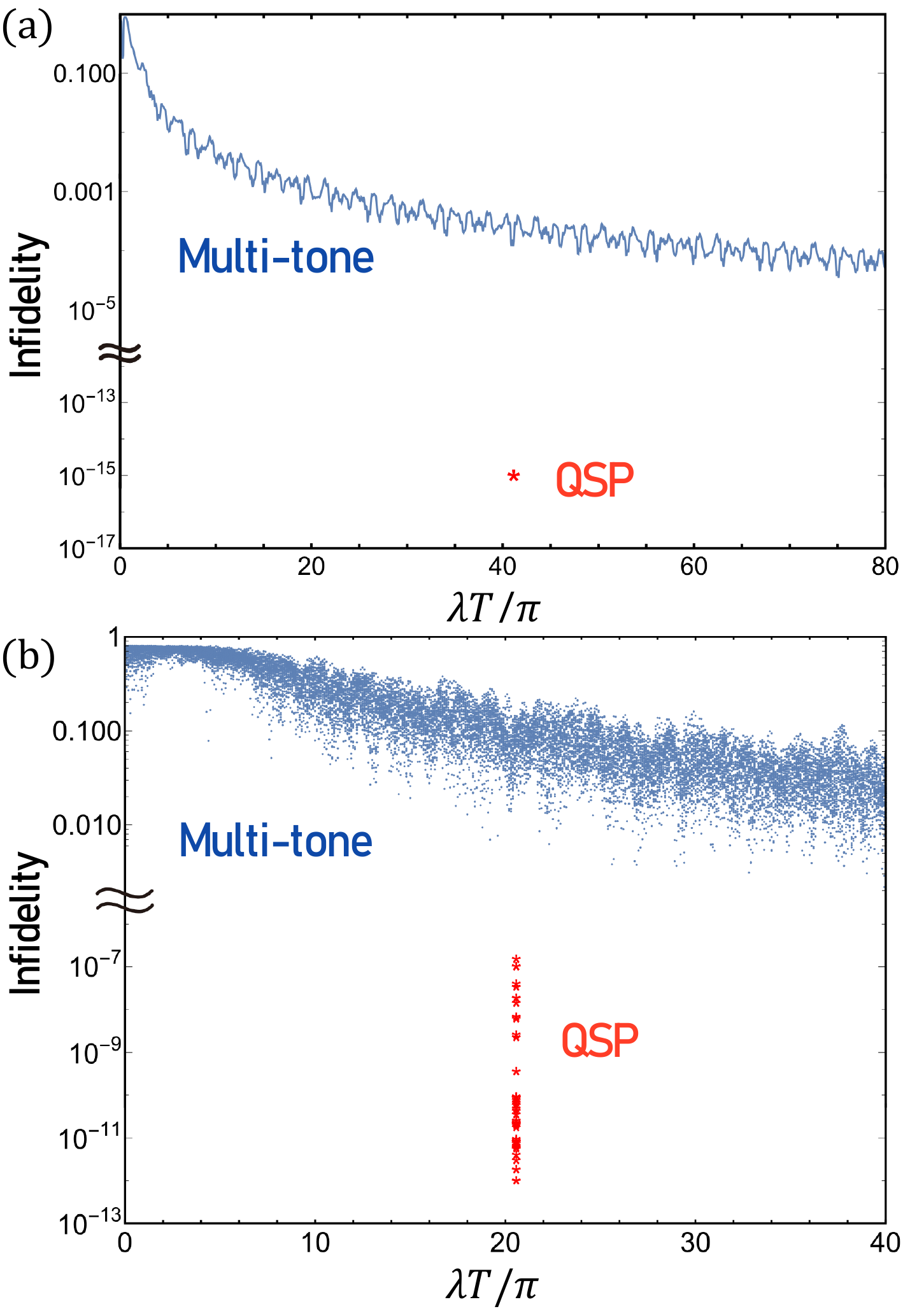}
    \caption{Gate infidelities of both the multi-tone driving (blue curve and dots) and QSP (red dots) implementations of SNAP gates with $N_{\text{max}}=3$ in a JC system. The details of the fidelities and the simulations can be found in Appendix~\ref{appendix_gate_fidelity} and \ref{appendix_multi} respectively. (a) Infidelity of the hybridization process that is described in Sec.~\ref{hybridization}. The QSP implementation achieves an infidelity that is eleven orders of magnitude lower under the same total implementation time, $T_{\text{total}} = 92\pi/(\sqrt{5}\lambda)$. 
    (b) Infidelity of the phase modification process that is described in Sec.~\ref{sec_JC_kernel}. The presented data consists of $50$ samples of the phase modification for SNAP gates with randomly chosen target phases. Our QSP method can generate gates with a fixed implementation time $ T_{\text{total}} = 46\pi/(\sqrt{5}\lambda)$. Their infidelities are orders of magnitude lower than those for multi-tone driving.}
    \label{fig_JC}
\end{figure}

We compare the performance of our QSP method in engineering the SNAP gate with the multi-tone driving method, which is detailed in Appendix~\ref{appendix_multi}. For the hybridization step, as shown in Fig.~\ref{fig_JC} (a), our QSP method achieves a gate infidelity that is eleven orders of magnitude lower than that of the multi-tone driving method under the same implementation time. Similarly, for the phase modification, Fig.~\ref{fig_JC} (a) shows that the QSP method consistently outperforms the multi-tone approach by several orders of magnitude in accuracy under the same implementation time.

Finally, we note that the process is very similar when the available interaction is anti-JC instead of JC, see the Appendix~\ref{appendix_aJC}.

\section{Non-unitary Operations}\label{sec_non_unitary}

In addition to unitary operations, our QSP method can also be extended to engineer non-unitary operations, which are irreversible and do not preserve the total probability. In this section, we present two illustrative examples: NLA and generalized-parity measurement.

\begin{figure}
    \centering
    \includegraphics[scale=0.18]{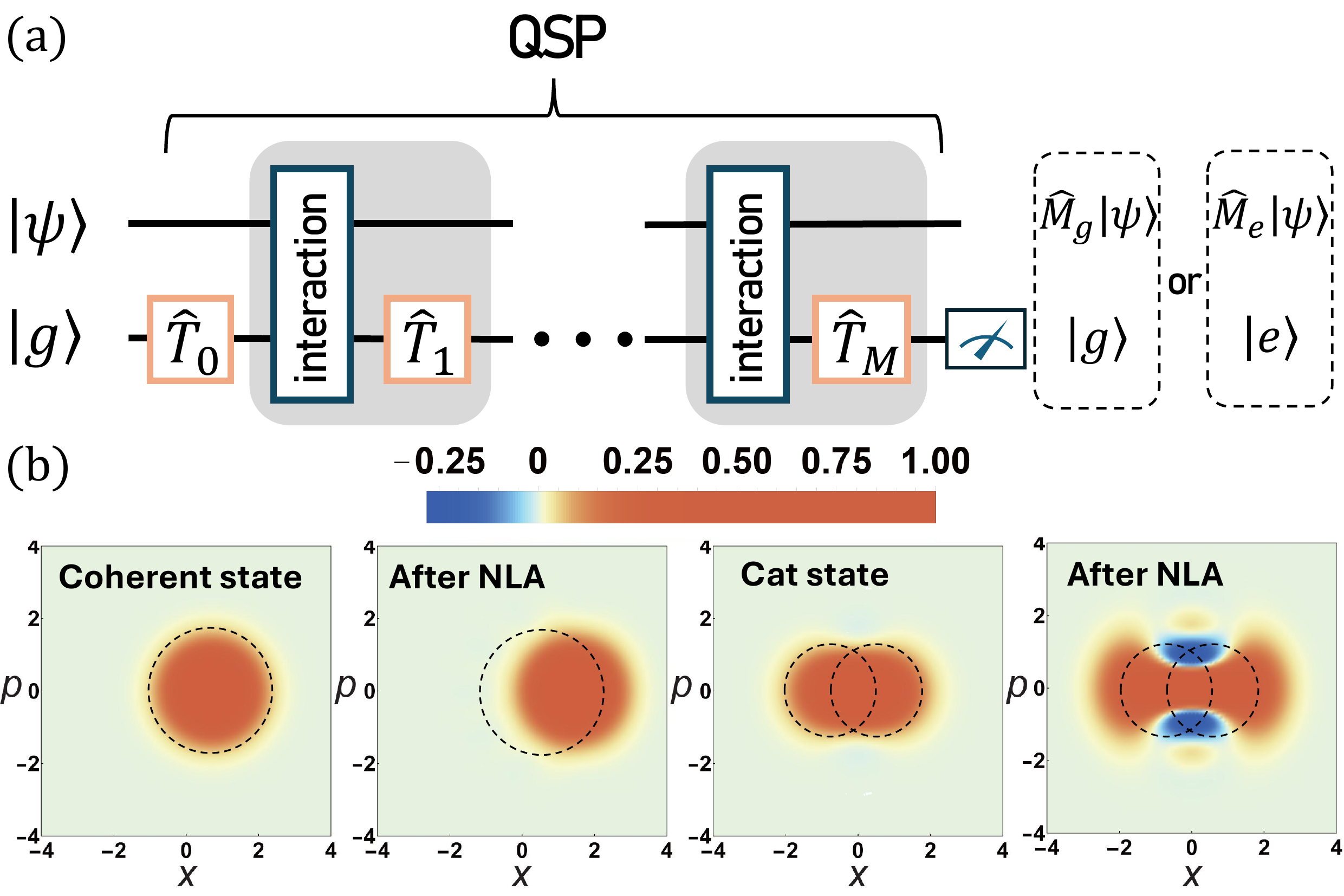}
    \caption{(a) Quantum circuit for engineering a non-unitary operation. Through QSP, the qumode and the qubit are entangled in a controllable way. After the qubit is measured, the qumode state $\ket{\psi}$ is non-unitarily transformed to either $\hat{M}_g\ket{\psi}$ or $\hat{M}_e\ket{\psi}$, depending on the qubit measurement outcome. (b) Wigner functions of a coherent state $\ket{\alpha}$ and a cat state $\ket{\alpha}+\ket{-\alpha}$ before and after NLA. The cat state has increased Wigner negativity after NLA. All initial states have amplitude $\alpha=0.5$ and are indicated by the dashed circles. They are amplified by an NLA with $G=2$ and $N_{\text{max}}=7$. } 
    \label{fig_circuit_non}
\end{figure}

Unlike gate engineering, which the qubit must be disentangled from the qumode afterwards, engineering non-unitary processes requires the qubits to be controllably entangled with the qumode. The desired transformation is then realized by performing a projective measurement on the qubit, as illustrated in Fig.~\ref{fig_circuit_non}(a). The protocol begins by entangling the qumode and qubit through the transformation
\begin{eqnarray}\label{non_U_formula}
    \ket{n}\ket{g} \to  A_n \ket{n}\ket{g} + B_n\ket{n}\ket{e} ,
\end{eqnarray}
where the amplitudes $A_n$ and $B_n$ satisfy the relation $|A_n|^2+|B_n|^2=1$. Subsequently, the qubit is measured in the computational basis $\{\ket{g},\ket{e}\}$. Depending on the outcome, the qumode undergoes a conditional, non-unitary transformation:
\begin{eqnarray}
 &&\text{Outcome}\,\ket{g}: \nonumber\\ 
 &&\frac{\hat{M}_g\ket{\psi}}{\sqrt{\bra{\psi}\hat{M}_g^\dag \hat{M}_g\ket{\psi}}} = \frac{1}{\sqrt{P_{g}}}\sum_{n=0}^{\infty} A_n c_n \ket{n},  \nonumber\\
 &&\text{Outcome}\,\ket{e}: \nonumber\\
 &&\frac{\hat{M}_e \ket{\psi}}{\sqrt{\bra{\psi}\hat{M}_e^\dag\hat{M}_e\ket{\psi}}} = \frac{1}{\sqrt{P_{e}}}\sum_{n=0}^{\infty} B_n c_n \ket{n}.\label{non_U_target_e}
\end{eqnarray}
Here $\ket{\psi}=\sum_{n=0}^\infty c_n\ket{n}$ is an arbitrary qumode state and the normalization factors correspond to the probabilities of each outcome: $P_{g} = \sum_{n=0}^{\infty} |A_n c_n|^2$ and $P_{e} = \sum_{n=0}^{\infty} |B_n c_n|^2$. The associated Kraus operators governing these transformations are
\begin{eqnarray}\label{kraus}
    \hat{M}_g  \equiv \sum_{n=0}^{\infty} A_n\ket{n}\bra{n} \quad \text{and} \quad
    \hat{M}_e \equiv  \sum_{n=0}^{\infty} B_n\ket{n}\bra{n}.\nonumber\\
\end{eqnarray}
In practice, the qumode state is usually dominated by a finite number of excitations, $N_{\text{max}}$, so we consider $n \in [0,N_{\text{max}}]$.

To engineer the entanglement process~\eqref{non_U_formula} in dispersive coupling systems, we should generally implement QSP with the polynomials satisfying $P(\omega_k^n)=A_n$ and $Q(\omega_k^n)=B_n$. We can construct the polynomial $P(z)$ by using the kernel method~\eqref{interpolation} with the following modification:
\begin{eqnarray}
    P(z) &=& \sum_{n=0}^{N_{\text{max}}} A_n K_n(z).
\end{eqnarray}
As mentioned earlier, the existing algorithm \cite{GQSP2024} can construct a polynomial that has the same norm as $Q(z)$ but with an undesired extra phase. The correction to the phase can be executed by a SNAP gate, which can also be engineered via QSP.

Similarly, for JC interaction systems, we design polynomials $F(z)$ and $G(z)$ to satisfy $F(e^{i\Phi_n})=e^{i\Upsilon_n}A_n$ and $G(e^{i\Phi_n})=e^{i\Upsilon_n}B_n$. This can be done by the kernel method~\eqref{kernel_modified} with the modification:
\begin{eqnarray}
    F(z) &=& \sum_{n=0}^{N_{\text{max}}}\Big[ \text{Re}\left[A_n \right]\mathcal{K}^R_n(z) + i\text{Im}\left[A_n\right] \mathcal{K}^I_n(z)\Big],\nonumber\\
\end{eqnarray}
the algorithm in Ref.~\cite{chao2020findinganglesquantumsignal} and the phase correction by a SNAP gate for $G(z)$.

\subsection{Quantum Noiseless Linear Amplification}

Signal amplification plays an important role in many quantum technologies, because quantum signals are usually too weak for accurate measurements \cite{Haus1962, caves1982, RMP_amp2010}. When the phase of a bosonic signal is unknown, phase-insensitive amplifications should be employed, but it inevitably adds noise due to the quantum uncertainty principle \cite{Haus1962, caves1982}.

NLA is an alternative amplification technique that can bypass this noise addition, but with a trade-off of success probability \cite{NLA_2009}. Specifically, it amplifies coherent states without increasing the uncertainty, i.e. it transforms any coherent state as $\ket{\alpha} \to \ket{G\alpha}$, where $G>1$ is the gain. For a general qumode state $\ket{\psi}$ with maximum excitation $N_{\text{max}}$, NLA transforms the state as
\begin{eqnarray}\label{NLA}
   \sum_{n=0}^{N_{\text{max}}} c_n\ket{n} \to \frac{1}{\sqrt{\mathcal{N}}}\sum_{n=0}^{N_{\text{max}}} G^{n} c_n\ket{n},
\end{eqnarray}
where the normalization factor is $\mathcal{N}=\sum_{n=0}^{N_{\text{max}}}G^{2n} |c_n|^2$. Fig.~\ref{fig_circuit_non} (b) illustrates the action of NLA on a coherent state and an even cat state.

Fundamental principles, including the no-cloning theorem \cite{Wootters1982} and the distinguishability of quantum states \cite{cloning_1998}, imply that NLA must be probabilistic. The optimal success probability is known to be proportional to $G^{-2N_{\text{max}}}$ \cite{NLA_limit_2013}. Despite this constraint, NLA has been applied to overcome channel loss \cite{Chrzanowski2014}, implement entanglement distillation \cite{Xiang2010}, enhance teleportation fidelity \cite{NLA_tele_2022} and improve the sensitivity of quantum illumination \cite{Karsa_2022}.
Several schemes for the implementation of NLA have been proposed and experimentally demonstrated \cite{Chrzanowski2014, Xiang2010, NLA_2020, NLA_2021, NLA_tele_2022, NLA_2022, NLA_cat_2023, NLA_PRX}. However, the success probability of most implementations is sub-optimal. Although NLA implementations with optimal probability have been proposed, they require either complex ancillary states \cite{NLA_2021, NLA_2022, NLA_cat_2023, NLA_PRX} or an interaction that is not commonly found in hybrid quantum systems \cite{NLA_arch_2014}.

By comparing transformations~\eqref{non_U_target_e} and \eqref{NLA}, we recognize that the optimal NLA can be realized with common hybrid interactions by using QSP. Our strategy is to engineer the amplitudes as $A_n = G^{n-N_{\text{max}}}$. NLA succeeds when the qubit is projected to $\ket{g}$. The success probability of our scheme, $P_{g}= G^{-2N_{\max}}\sum_{n=0}^{N_{\text{max}}} G^{2n}|c_n|^2$, achieves the theoretical optimum \cite{NLA_arch_2014}. 

\subsection{Generalized-Parity Measurement}

Boson-number and parity measurements are useful in many tasks, such as generating non-Gaussian resources \cite{PhysRevLett.97.110501, PhysRevA.85.062318, zeytinoglu2024}, detecting error-syndrome in bosonic codes \cite{Sun2014}, and bosonic state tomography \cite{Kirchmair2013, PhysRevA.94.052327, PRXQuantum.6.010303}.
Here, we propose using QSP to realize a generalized-parity measurement, which can serve as a boson-number measurement under suitable parameter choices. Specifically, this measurement determines whether the boson number $n$ satisfies the modular relation $n \,(\text{mod}\, k') = n_b$ for any integer modulus $k'$ of choice. The associated projective operators for positive and negative outcomes are, respectively,
\begin{eqnarray}\label{project}
    \hat{M}_{n_b} &\equiv& \sum_{m=0}^\infty \ket{mk'+n_b}\bra{mk'+n_b}, \quad \text{and}\nonumber\\
    \hat{N}_{n_b}&\equiv& \hat{I}-\hat{M}_{n_b}.
\end{eqnarray}
This reduces to the standard parity measurement when choosing $k'=2$ and $n_b=0$.

Comparing the operators in Eq.~\eqref{project} with the Kraus operators~\eqref{kraus}, it is clear that they can be engineered via QSP by choosing $A_n=0$ and $|B_n|=1$ for $n \,(\text{mod}\, k') = n_b$, and $A_n=1$ and $B_n=0$ otherwise.
Upon measuring on the auxiliary qubit, the outcome $\ket{e}$ heralds the modular relation being satisfied, and the outcome $\ket{g}$ signals otherwise. Notably, this measurement is particularly suitable to implement on the systems with dispersive coupling due to its natural compatibility with modular arithmetic.

A complete boson-number measurement is achieved by sequentially applying generalized-parity measurements with a fixed $k'=N_{\text{max}}+1$ and a varying $n_b$. Specifically, one can begin with an arbitrary $n_b$. If the measurement yields the positive projection $\hat{M}_{n_b}$, the boson number is identified. Otherwise, one conducts measurement with different $n_b$ until a positive projection is obtained.

During the preparation of this manuscript, we became aware of Ref.~\cite{zeytinoglu2024}, which also proposes using QSP to realize generalized-parity measurements. However, their approach generates merely an approximation of the desired projection operators and is applicable to only dispersive coupling systems. By contrast, our scheme provides an exact implementation and is compatible with both dispersive coupling and JC systems.

\section{Conclusion}\label{sec_conclusion}

In this work, we leverage the techniques in QSP to engineer both unitary and non-unitary operations in hybrid qumode-qubit systems. Our approach can generate accurate non-Gaussian operations and can be implemented with diverse types of hybrid interaction, including dispersive and JC couplings.

In dispersively coupled systems, our QSP method can engineer a new type of non-Gaussian gate, the mod-$k$ gate. It is a generalization of the eParity gate and can be reduced to the SNAP gate with a suitable choice of parameters. Remarkably, the mod-$k$ gate can be engineered in a fixed time $4\pi/\chi$ with high accuracy and independence on the number of levels involved. Due to the unique features of this gate, it has a variety of potential DV and CV applications, such as acting as logic gates for bosonic codes and deterministic generation of multiple-component cat states.
Furthermore, we extend the mod-$k$ gate to two modes and illustrate its capability to implement generic entangling gates for qudits and bosonic codes.

Thanks to QSP's capability to controllably establish qumode-qubit entanglement, a wide range of desired non-unitary transformations can also be engineered by our QSP method. We discussed the implementations of several non-unitary operations with practical relevance, including NLA with optimal success probability and generalized-parity measurements.

Lastly, our generic kernel method offers a streamlined and versatile approach to designing QSP polynomials. It could find applications in other algorithms involving QSP, such as QSP interferometry \cite{Sinanan_Singh_2024}, Hamiltonian simulation \cite{crane2024}, and state transfer in hybrid qumode-qubit systems \cite{liu2024}. Our results open new avenues for advancing bosonic quantum technologies.


\begin{acknowledgments}
This work is supported by the Natural Sciences and Engineering Research Council of Canada Discovery Grant (NSERC RGPIN-2021-02637), Alliance International Catalyst Quantum Grant (ALLRP 578638-22), and Canada Research Chairs (CRC-2020-00134).
\end{acknowledgments}

\section*{data availability}
The source code supporting the numerical results in this article is publicly available at Zenodo \cite{Fong2026QSPCode}.

\appendix

\section{QSP Conventions}\label{appendix_convention}

In literature \cite{QSP2016, QSP2017, QSP2019, QSP2021, GQSP2024}, (original) QSP is formulated as a sequence of alternative signal operators and signal-processing (hereafter, processing) operators. The specific unitaries represented by these operators depend on the literature. In this appendix, we briefly review two conventional representations of QSP in the literature. We will discuss two major conventions, Wz and Wx, which were discussed in Ref.~\cite{QSP2021}. We note that in this work, our oQSP formalism follows a variant of the Wz convention, originally introduced in Refs.~\cite{QSP2021} and \cite{chao2020findinganglesquantumsignal}.

\subsection{Wz Convention}\label{app_wz}

In this convention, the signal operation is the phase-shift defined as
\begin{eqnarray}
    \hat{W}_\Phi \equiv e^{-i\Phi/2} \hat{Z}_\Phi = \begin{pmatrix}
        e^{i\Phi/2} & 0\\
        0 & e^{-i\Phi/2}
    \end{pmatrix}.
\end{eqnarray}
The processing operation is the rotation about $x$-axis, namely $\hat{R}_{x,m}\equiv \hat{R}_m$ (c.f. Eq.~\eqref{rotation_matrix}) with $\phi_m=\lambda_m=\pi/2$ and arbitrary $\theta_m$. After $M$ rounds of signal and processing operators, the overall unitary is
\begin{eqnarray}\label{QSP_sequence_appendix}
    \hat{U}_{\text{oQSP,z}} &=& \prod_{m=1}^{M} \left(\hat{R}_{x,m}\hat{W}_\Phi\right) \hat{R}_{x,0} \nonumber\\
    &=&\begin{pmatrix}
\mathcal{F}(e^{i\Phi/2}) & \mathcal{G}(e^{-i\Phi/2}) \\
\mathcal{G}(e^{i\Phi/2}) & \mathcal{F}(e^{-i\Phi/2})
\end{pmatrix},
\end{eqnarray}
where $\mathcal{F}(w)$ and $\mathcal{G}(w)$ are degree-$M$ Laurent series in the complex variable $w\equiv e^{i\Phi/2}$:
\begin{eqnarray}\label{F_G_special}
    \mathcal{F}(w) =  \sum_{m=-M}^M \bar{f}_m w^{m} \quad \text{and} \quad
    \mathcal{G}(w) =  \sum_{m=-M}^M i\bar{g}_m w^{m},\nonumber\\
    \label{G_w}
\end{eqnarray}
with real coefficients $\bar{f}_m$ and $\bar{g}_m$. These coefficients are non-zero only for $m = 2l + (M \,\text{mod}\, 2)$, where $l$ is an integer.
By adjusting the rotation angles $\{\theta_m\}$, one can engineer these coefficients to take arbitrary real values under the constraint
\begin{eqnarray} 
    |\mathcal{F}(w)|^2+|\mathcal{G}(w)|^2 = 1,
\end{eqnarray}
for any $\Phi$.

Comparing Eqs.~\eqref{QSP_sequence_appendix} and \eqref{QSP_sequence_1}, we identify the relation between our polynomials $F(z)$ and $G(z)$ (c.f. \eqref{G_z}), with $z\equiv e^{i\Phi} =w^2$, and the Laurent series $\mathcal{F}(w)$ and $\mathcal{G}(w)$ (c.f. \eqref{G_w}) as
\begin{eqnarray}\label{F_to_F}
    F(w^2) = w^{M} \mathcal{F}(w) \quad \text{and} \quad G(w^2) = w^{M} \mathcal{G}(w).\nonumber\\
\end{eqnarray}
It follows that the polynomial coefficients are related by
\begin{eqnarray}
    f_{m} = \bar{f}_{2m-M} \quad \text{and} \quad g_{m} = \bar{g}_{2m-M},
\end{eqnarray}
for all $0\leq m\leq M$. Since $\bar{f}_{2m-M+1}=\bar{g}_{2m-M+1}=0$, they are not related to any $f_m$ and $g_m$. Furthermore, Eq.~\eqref{F_to_F} implies that when $\mathcal{F}(w)$ and $\mathcal{G}(w)$ are Laurent series of degree $M$ in $w$, $F(w^2)$ and $G(w^2)$ are polynomials of degree $2M$ in $w$, or equivalently, polynomials of degree $M$ in $z$.
 
\subsection{Wx Convention}\label{app_wx}

An alternative convention, introduced in Refs.~\cite{QSP2016, QSP2017}, takes the signal operators to be a rotation about $x$-axis: 
\begin{eqnarray}
    \hat{X}_\Phi &\equiv& \hat{R}_{H}\hat{W}_\Phi\hat{R}_{H} =\begin{pmatrix}
        a & i\sqrt{1-a^2}\\
        i\sqrt{1-a^2} & a
    \end{pmatrix},\nonumber\\
\end{eqnarray}
where $a\equiv\cos(\Phi/2)$ and $\hat{R}_{H}$ denotes the Hadamard rotation, i.e. the rotation with $\phi=\lambda=0$ and $\theta=\pi/4$. This is referred to as the Wx convention in Ref.~\cite{QSP2021}.

In this convention, the processing operators are rotations about the $z$-axis, namely $\hat{R}_{z,m}\equiv\hat{R}_H \hat{R}_{x,m}\hat{R}_H$. After $M$ rounds of signal operators and processing operators, the overall operator becomes
\begin{eqnarray}
    \hat{U}_{\text{oQSP,x}} &=& \prod_{m=1}^M \left(\hat{R}_{z,m}\hat{X}_\Phi\right)\hat{R}_{z,0} \nonumber\\
    &=& 
    \begin{pmatrix}
     \mathcal{R}(a) & i\mathcal{T}(a)\sqrt{1-a^2}\\
     i\mathcal{T}^*(a)\sqrt{1-a^2} & \mathcal{R}^*(a)
    \end{pmatrix}.\nonumber\\
\end{eqnarray}
Here, $\mathcal{R}(a)$ and $\mathcal{T}(a)$ are polynomials that can be expanded in the Chebyshev bases \cite{mason2002chebyshev},
\begin{eqnarray}\label{chebyshev}
    \mathcal{R}(a) &=& \sum_{m=0}^M r_m T_m(a),\nonumber\\
    \mathcal{T}(a) &=& \sum_{m=1}^M t_m U_{m-1}(a),
\end{eqnarray}
where $T_m(a)$ and $U_{m}(a)$ are the degree-$m$ Chebyshev polynomials of the first kind and the second kind, respectively. By varying the rotations $\{\theta_m\}$, the coefficients $r_k$ and $t_k$ can be engineered to take arbitrary complex values under the constraint
\begin{eqnarray}
    |\mathcal{R}(a)|^2+(1-a^2)|\mathcal{T}(a)|^2=1,
\end{eqnarray}
for any $-1\leq a\leq 1$.

We note that the polynomials in Wx convention differ from those in the convention of this manuscript in both variables and structure. In Wx convention, polynomials $\mathcal{R}(a)$ and $\mathcal{T}(a)$ are functions of $a\equiv \cos (\Phi/2)$ and have different degrees, as shown in Eq.~\eqref{chebyshev}: when $\mathcal{R}(a)$ is a degree-$M$ polynomial, $\mathcal{T}(a)$ is a degree-$(M-1)$ polynomial. In contrast, in our convention the polynomials $F(z)$ and $G(z)$ are functions of $z\equiv e^{i\Phi}$ and have the same degree.

The Wx convention is related to the Wz conventions through a Hadamard conjugation:
\begin{eqnarray}
    \hat{U}_{\text{oQSP,x}} &=& \prod_{m=1}^M \left(\hat{R}_{z,m}\hat{X}_\Phi\right)\hat{R}_{z,0} \nonumber\\
    &=& 
\hat{R}_{H}\prod_{m=1}^M \left(\hat{R}_{x,m}\hat{W}_\Phi\right)\hat{R}_{x,0}\hat{R}_{H} \nonumber\\
    &=& \hat{R}_{H} \hat{U}_{\text{oQSP,z}}\hat{R}_{H}.
\end{eqnarray}
This relation allows one to map the polynomial coefficients between two conventions, i.e.
\begin{eqnarray}
    \bar{f}_0 = \text{Re}[r_0] \quad \text{and}\quad \quad \bar{g}_0 = -i\text{Im}[r_0],
\end{eqnarray}
and
\begin{eqnarray}
    &&\bar{f}_m = \frac{1}{2}\text{Re}[r_m+t_m], \quad \bar{g}_m = -\frac{i}{2}\text{Im}[r_m-t_m], \nonumber\\
    &&\bar{f}_{-m} = \frac{1}{2}\text{Re}[r_m-t_m], \quad \bar{g}_{-m} = -\frac{i}{2}\text{Im}[r_m+t_m],\nonumber\\
\end{eqnarray}
for all $0<m \leq M$.

\section{Multi-level QSP in Eigenstate Basis}\label{appendix_big_matrix}

As presented in Ref.~\cite{GQSP2024}, QSP can also be applied on a $N$-level {\it system} that couples to an auxiliary {\it control qubit}. In this setting, the signal and processing operators are $2N\times2N$ matrices. Generally, we can express the operators in the basis $\{\ket{\psi_{0}}\ket{g},...,\ket{\psi_{N-1}}\ket{g}, \ket{\psi_{0}}\ket{e},...,\ket{\psi_{N-1}}\ket{e}\}$, where $\ket{g}$ and $\ket{e}$ are the qubit basis state and $\{\ket{\psi_n}\}$ is an arbitrary basis of the system. In this appendix, we will use GQSP as an example to show that the formalism of the multi-level QSP is equivalent to our two-level model when one considers the eigenstates of the multi-level system. 

Particularly, the signal operator of GQSP is a controlled evolution, where the system evolves under a Hamiltonian when the qubit is in $\ket{g}$ and remains unchanged when the qubit is in $\ket{e}$:
\begin{eqnarray}
    {\bf Z}_{\bf H} =
    \begin{pmatrix}
        e^{i{\bf H}} & 0 \\
        0 & \mathbb{I}
    \end{pmatrix},
\end{eqnarray}
where ${\bf H}$ represents an $N\times N$ hermitian matrix that characterizes the evolution and $\mathbb{I}$ is the $N\times N$ identity matrix.

The processing operator is implemented as a qubit rotation, so it is an identity to the system. The corresponding matrix is given by
\begin{eqnarray}\label{R_big}
    {\bf R}_m =
    \begin{pmatrix}
        e^{i(\lambda_m+\phi_m)}\cos\theta_m \, \mathbb{I}
        &
        e^{i\phi_m}\sin\theta_m \, \mathbb{I}
        \\
        e^{i\lambda_m}\sin\theta_m \, \mathbb{I}
        &
        -\cos\theta_m \, \mathbb{I}
    \end{pmatrix},\nonumber\\
\end{eqnarray}
The overall matrix expression of the $M$-step GQSP sequence is then
\begin{eqnarray}\label{eq_app_qsp}
    {\bf U}_{\text{GQSP}}
    =
    \prod^M_{m=1}
    \left(
        {\bf R}_m {\bf Z}_{\bf H}
    \right)
    {\bf R}_0
    =
    \begin{pmatrix}
        P(e^{i{\bf H}}) & \boldsymbol{\cdot} \\
        Q(e^{i{\bf H}}) & \boldsymbol{\cdot}
    \end{pmatrix}.\nonumber\\
\end{eqnarray}

Although this expression appears to describe a genuinely $2N$-level transformation, its underlying structure can be understood as a collection of independent two-level QSP evolutions. This becomes transparent by representing the system in the eigenstate basis of ${\bf H}$, $\{\ket{\Phi_0},...,\ket{\Phi_{N-1}}\}$, which satisfy ${\bf H}\ket{\Phi_n}=\Phi_n\ket{\Phi_n}$. In this basis, the matrix representation of the signal operator become diagonal,
\begin{eqnarray}
    {\bf Z}_{\bf H} &=& 
    \begin{pmatrix}
        e^{i{\bf \Phi}} & 0 \\
        0 & \mathbb{I}
    \end{pmatrix},
\end{eqnarray}
where $ {\bf \Phi}= \text{diag}(\Phi_0,\Phi_1,...,\Phi_{N-1})$. Since each ${\bf R}_m$ acts only on the control qubit, its expression remains the same for any basis choice of the system. Therefore, in the eigenbasis of ${\bf H}$, the full GQSP sequence becomes
\begin{eqnarray}\label{gqsp_app}
    {\bf U}_{\text{GQSP}} =
    \prod^M_{m=1}
    \left[
        {\bf R}_m
        \begin{pmatrix}
            e^{i{\bf \Phi}} & 0 \\
            0 & \mathbb{I}
        \end{pmatrix}
    \right]
    {\bf R}_0.
\end{eqnarray}

In this representation, both the signal and processing operators are represented by a matrix with four $N\times N$ diagonal blocks. As a consequence, any state $\ket{\Phi_n}\ket{g}$ can be transformed to only $\ket{\Phi_n}\ket{e}$, and vice versa. These correspond to the dressed-state basis $\ket{\downarrow_n}=\ket{\Phi_n}\ket{g}$ and $\ket{\uparrow_n}=\ket{\Phi_{n}}\ket{e}$ in the main text. Each $\{\ket{\Phi_n}\ket{g},\ket{\Phi_n}\ket{e}\}$ subspace undergoes an independent two-level QSP transformation  with phase parameter $\Phi_n$. The overall $2N$-level transformation is therefore equivalent to a collection of independent transformation of $N$ decoupled two-level systems.

The same principle also applies to oQSP that involves a control qubit and a multi-level system. In the eigenbasis of the signal operator, both signal and processing operators consist of four $N\times N$ diagonal blocks. Therefore, QSP can be interpreted as applying transformations in parallel across mutliple subspaces, each spanned by two dressed states.

\section{Kernel Method for Dispersive Coupling Systems}\label{appendix_bound}

To engineer both unitary and non-unitary operations in dispersive coupling systems, we construct the polynomial $P(z)$ via the kernel method:
\begin{eqnarray}\label{int_app}
    P(z) =\sum_{n=0}^{k-1} A_n K_n(z),
\end{eqnarray}
where $A_n$ are arbitrary complex target amplitudes that are bounded by unity, i.e. $|A_n|\leq 1$, and the kernels are given by Eqs.~\eqref{kn_k0} and \eqref{kernel_form}. By construction, they satisfy the necessary condition $K_n(\omega_k^m)=\delta_{nm}$. In this appendix, we prove that they also allow the normalization requirement $|P(z)| \leq 1$ to be satisfied for any set of target amplitudes.

Starting from Eq.~\eqref{int_app}, we consider the triangle inequality to bound its absolute value:
\begin{eqnarray}\label{bound}
|P(z)| = \left|\sum_{n=0}^{k-1}A_n K_n(z)\right| 
&\leq& \sum_{n=0}^{k-1} \left|A_n\right| \left|K_n(z)\right|\nonumber\\
&\leq& \sum_{n=0}^{k-1} \left|K_n(z)\right|.
\end{eqnarray}
Using the property~\eqref{kn_k0}, the inequality becomes 
\begin{eqnarray}\label{P_bound}
    |P(z)| \leq \sum_{n=0}^{k-1} \left|K_0(z \omega_k^{-n})\right|.
\end{eqnarray}
To determine the bound of the summation in Eq.~\eqref{P_bound}, we first write the absolute square of Eq.~\eqref{kernel_form} as
\begin{eqnarray}\label{rewrite_k}
        |K_0(z)|^2 &=& \frac{1}{16k^4} \left(z+1\right)^2\left(z^*+1\right)^2\nonumber \\
        &\times& \prod_{m=1}^{k-1}\left(z-\omega_k^{m}\right)^2\left(z^*-\omega_k^{-m}\right)^2,
\end{eqnarray}
where we apply the identity $\prod^{k-1}_{m=1}(1-\omega_k^m)=k$ to simplify the denominator.
Then, the square root of Eq.~\eqref{rewrite_k} can be easily found as
\begin{eqnarray}
     |K_0(z)| &=& \frac{1}{4k^2} \left(z+1\right)\left(z^*+1\right)\nonumber \\
        &\times& \prod_{m=1}^{k-1}\left(z-\omega_k^{m}\right)\left(z^*-\omega_k^{-m}\right).\label{product_kernel}
\end{eqnarray}

Since we are considering the value of $|P(z)|$ on the unit circle $|z|=1$, we can rewrite $z^*=z^{-1}$ and hence Eq.~\eqref{product_kernel} can be expressed as a Laurent series $|K_0(z)| = \sum_{m=-k}^{k} \beta_m z^m$, where $\beta_m$ are complex coefficients. Then, the summation in Eq.~\eqref{P_bound} can be rewritten as 
\begin{eqnarray}
    \sum_{n=0}^{k-1} \left|K_0(z \omega_k^{-n})\right|=\sum_{m=-k}^k \left(\sum_{n=0}^{k-1} \omega_k^{-nm}\right) \beta_m z^m.\nonumber\\
\end{eqnarray}
Since the sum over $k$th roots of unity satisfies $\sum_{n=0}^{k-1} \omega_k^{-nm} = k(\delta_{m,k}+\delta_{m,0}+\delta_{m,-k})$, it follows that 
\begin{eqnarray}\label{sum_k}
   \sum_{n=0}^{k-1} \left|K_0(z \omega_k^{-n})\right| 
&=& k(\beta_{k} z^k + \beta_{0} + \beta_{-k} z^{-k}) .
\end{eqnarray}
Direct expanding Eq.~\eqref{product_kernel} gives the coefficients
   $ \beta_k = \beta_{-k} = \frac{1}{4k^2}$ and 
    $\beta_0  = \frac{1}{k}-\frac{1}{2k^2}$. 
Since these coefficients are real, the upper bound of Eq.~\eqref{sum_k} is given by $ k(\beta_{k} + \beta_{0} + \beta_{-k}) =1$. 
Hence, $|P(z)|\leq \sum_{n=0}^{k-1} \left|K_0(z \omega_k^{-n})\right|\leq 1$, which proves the normalization condition. 

\section{Examples of Gate Engineering with QSP}\label{appendix_gate_example}

In addition to the eParity gate in Sec.~\ref{sec_ep}, we present two more examples of gate engineered by QSP in this appendix. 

\subsection{SNAP Gates for ground state $\ket{0}$ component}

SNAP gate is a commonly employed non-Gaussian gate in CV QIP. By imprinting independently tunable phases on individual Fock states, the SNAP gate provides a versatile tool for manipulating oscillator states beyond the Gaussian regime. As a simple illustration, we consider a SNAP gate that modifies only the phase of the ground state $\ket{0}$ component.

The target SNAP gate is defined by Eq.~\eqref{equation_snap}, with 
\begin{eqnarray}
    \Theta_n=\begin{cases}
        \Theta_0 \quad \text{for}\quad n=0\\
        0 \quad  \text{for} \quad n=1,...,N_{\text{max}}
    \end{cases}.
\end{eqnarray}
This gate can be realized using the mod-$k$ gate with $k=N_{\text{max}}+1$ and $\Theta_{n\, (\text{mod}\, k)} = \Theta_0 \delta_{n0}$. To engineer this mod-$k$ gate, we first construct the polynomial $P(z)$ using the kernel method in Eq.~\eqref{interpolation}. The complementary polynomial $Q(z)$ is then obtained by solving Eq.~\eqref{PQ_constraint}, i.e. $|Q(z)|^2=1-|P(z)|^2$. 

As a numerical example, we consider the case $k=5$ with the target phase $\Theta_0=\pi$; this gate is useful for preparing Fock states \cite{SNAP_2015}. The coefficients $p_m$ and $q_m$ of the corresponding polynomials are listed in the table~\ref{tab_of_snap_gate1}. Once these polynomial coefficients are determined, the rotation angles can be obtained using iteratively using the algorithm of Ref.~\cite{GQSP2024}. The corresponding results are also shown in table~\ref{tab_of_snap_gate1}. For this example, the gate infidelity is of the order of $10^{-15}$, which is significantly smaller than that obtained using the multi-tone driving method for the same operation time $\chi T=4\pi$, where the infidelity is of order $10^{-2}$. 


\begin{table}[h!]
    \centering
    For SNAP gate with target phase $\Theta_0=\pi$
    \begin{tabularx}{0.45\textwidth}{|>{\centering\arraybackslash}X|>{\centering\arraybackslash}X|>{\centering\arraybackslash}X|>{\centering\arraybackslash}X|}
    \hline
    \multicolumn{2}{|c|}{$P(z)$}    & \multicolumn{2}{|c|}{$Q(z)$}     \\
      \hline
      $p_0 $   &  $0.03$ & $q_0$ & $0.03047 + 0.08154 i$   \\
      \hline
        $p_1$   &  $-0.02472 + 0.07608 i$ & $q_1$ & $-0.3425 + 0.0149 i$   \\
        \hline
        $p_2$   &  $0.1294 + 0.09405 i$ & $q_2$ & $-0.04748 + 0.1711i$ \\
       \hline
       $p_3$   &  $0.1942 - 0.1411 i$ & $q_3$ & $0.06043 + 0.04001 i$  \\
       \hline
       $p_4$   &  $-0.09889 - 0.3043 i$ & $q_4$ & $0.01129 - 0.008977 i$  \\
       \hline
       $p_5$   &  $0.54$ & $q_5$ & $-0.02685 - 0.07185 i$  \\
       \hline
       $p_6$   &  $-0.09889 + 0.3043i$ & $q_6$ & $0.3425 - 0.0149i$  \\
       \hline     
       $p_7$   &  $0.1942 + 0.1411 i$ & $q_7$ & $0.04748 - 0.1711i$  \\
       \hline  
       $p_8$   &  $0.1294 - 0.09405 i$ & $q_8$ & $-0.06043 - 0.04001 i$  \\
       \hline  
       $p_9$   &  $-0.02472 - 0.07608 i$ & $q_9$ & $-0.01129 + 0.008977 i$  \\
       \hline  
       $p_{10}$   &  $0.03$ & $q_{10}$ & $-0.00362 - 0.009685 i$  \\
       \hline         
    \end{tabularx}
Rotation angles
    \begin{tabularx}{0.45\textwidth}{|>{\centering\arraybackslash}X|>{\centering\arraybackslash}X|}
    \hline
      $\{\theta_0,\phi_0,\lambda_0\}$   &  $\{\theta_1,\phi_1,\lambda_1\}$   \\
      \hline
        $\{0.394\pi, 1.6\pi, 0.386\pi\}$ &  $\{0.119\pi, 0.6\pi, 0\}$   \\
        \hline
        $\{\theta_2,\phi_2,\lambda_2\}$ &  $\{\theta_3,\phi_3,\lambda_3\}$  \\
       \hline
       $\{0.124\pi, 0.6\pi, 0\}$ &  $\{0.105\pi, 0.6\pi, 0\}$   \\
       \hline
           $\{\theta_4,\phi_4,\lambda_4\}$ & $\{\theta_5,\phi_5,\lambda_5\}$  \\
        \hline
        $\{0.0711\pi, 0.6\pi, 0\}$ &  $\{0.451\pi, 0.6\pi, 0\}$   \\
        \hline
           $\{\theta_6,\phi_6,\lambda_6\}$ & $\{\theta_7,\phi_7,\lambda_7\}$  \\
        \hline
        $\{0.0711\pi, 0.6\pi, 0\}$ &  $\{0.105\pi, 0.6\pi, 0\}$   \\
        \hline    
      $\{\theta_8,\phi_8,\lambda_8\}$ & $\{\theta_9,\phi_9,\lambda_9\}$  \\
        \hline
        $\{0.124\pi, 0.6\pi, 0\}$ &  $\{0.119\pi, 0.6\pi, 0\}$   \\
        \hline 
           $\{\theta_{10},\phi_{10},\lambda_{10}\}$ & $/$  \\
        \hline
        $\{0.106\pi, 0.614\pi, 0\}$ &  $/$   \\
        \hline         
    \end{tabularx}
    \caption{Table of polynomial coefficients and corresponding rotation angles for engineering a SNAP gate with target phase $\Theta_0=\pi$ using QSP.}
        \label{tab_of_snap_gate1}
    \end{table}

\subsection{mod-$k$ Gates for cat states generation}

As discussed in Sec.~\ref{sec_app_cat}, the mod-$k$ gate can also be used for cat state generation. When the target phases satisfy Eq.~\eqref{relation}, the mod-$k$ gate is equivalent to an equal-amplitude superposition of $k$ bosonic rotations, Eq.~\eqref{superposition_of_rotation}. Therefore, it transforms a coherent state to an equal-amplitude $k$-component cat state, Eq.~\eqref{cat_eq_eq}.

For the example demonstrated in Fig.~\ref{fig_cat}, we choose $k=5$. The target phases are $\{ \Theta_{0\, (\text{mod}\, 5)},\Theta_{1\, (\text{mod}\, 5)},\Theta_{2\, (\text{mod}\, 5)},\Theta_{3\, (\text{mod}\, 5)},\Theta_{4\, (\text{mod}\, 5)} \} $ $= \{\pi,-\pi/5,\pi/5,\pi/5,-\pi/5 \}$. The coefficients $p_m$ and $q_m$ of corresponding polynomials are determined and shown in table~\ref{tab_of_cat}. The associated rotation angles are also given in the same table.

\begin{table}[h!]
    \centering
    mod-$5$ gate for $5$-component cat state
    \begin{tabularx}{0.45\textwidth}{|>{\centering\arraybackslash}X|>{\centering\arraybackslash}X|>{\centering\arraybackslash}X|>{\centering\arraybackslash}X|}
    \hline
    \multicolumn{2}{|c|}{$P(z)$}    & \multicolumn{2}{|c|}{$Q(z)$}     \\
      \hline
      $p_0 $   &  $0.02236$ & $q_0$ & $-0.03855 + 0.01433 i$   \\
      \hline
        $p_1$   &  $-0.07236+0.05257i$ & $q_1$ & $0.02665 - 0.1165 i$   \\
        \hline
        $p_2$   &  $0.1789$ & $q_2$ & $-0.3611 - 0.1727 i$ \\
       \hline
       $p_3$   &  $0.08291 - 0.2552 i$ & $q_3$ & $0.001583 - 0.02477 i$  \\
       \hline
       $p_4$   &  $0.1106 - 0.3403 i$ & $q_4$ & $0.02040 + 0.04073 i$  \\
       \hline
       $p_5$   &  $0.402492$ & $q_5$ & $0.02716 - 0.01009 i$  \\
       \hline
       $p_6$   &  $-0.2894 + 0.2103 i$ & $q_6$ & $-0.02665 + 0.1165 i$  \\
       \hline     
       $p_7$   &  $0.2683$ & $q_7$ & $0.3611 + 0.1727 i$  \\
       \hline  
       $p_8$   &  $0.05528 - 0.1701 i$ & $q_8$ & $-0.001583 + 0.02477 i$  \\
       \hline  
       $p_9$   &  $0.02764 - 0.08507 i$ & $q_9$ & $-0.02040 - 0.04073i$  \\
       \hline  
       $p_{10}$   &  $0.02236$ & $q_{10}$ & $0.01140 - 0.004236i$  \\
       \hline         
    \end{tabularx}
Rotation angles
    \begin{tabularx}{0.45\textwidth}{|>{\centering\arraybackslash}X|>{\centering\arraybackslash}X|}
    \hline
      $\{\theta_0,\phi_0,\lambda_0\}$   &  $\{\theta_1,\phi_1,\lambda_1\}$   \\
      \hline
        $\{0.342\pi, 0.144\pi, 0.887\pi\}$ &  $\{0.294\pi, 0.469\pi, 0\}$   \\
        \hline
        $\{\theta_2,\phi_2,\lambda_2\}$ &  $\{\theta_3,\phi_3,\lambda_3\}$  \\
       \hline
       $\{0.242\pi, 0.881\pi, 0\}$ &  $\{0.399\pi, 0.205\pi, 0\}$   \\
       \hline
           $\{\theta_4,\phi_4,\lambda_4\}$ & $\{\theta_5,\phi_5,\lambda_5\}$  \\
        \hline
        $\{0.192\pi, 0.766\pi, 0\}$ &  $\{0.0985\pi, 1.57\pi, 0\}$   \\
        \hline
           $\{\theta_6,\phi_6,\lambda_6\}$ & $\{\theta_7,\phi_7,\lambda_7\}$  \\
        \hline
        $\{0.109\pi, 0.0767\pi, 0\}$ &  $\{0.0914\pi, 1.59\pi, 0\}$   \\
        \hline    
      $\{\theta_8,\phi_8,\lambda_8\}$ & $\{\theta_9,\phi_9,\lambda_9\}$  \\
        \hline
        $\{0.265\pi, 1.29\pi, 0\}$ &  $\{0.192\pi, 0.0153\pi, 0\}$   \\
        \hline 
           $\{\theta_{10},\phi_{10},\lambda_{10}\}$ & $/$  \\
        \hline
        $\{0.159\pi, 0.113\pi, 0\}$ &  $/$   \\
        \hline         
    \end{tabularx}
    \caption{Table of polynomial coefficients and corresponding rotation angles for engineering a mod-$5$ gate using QSP.}
        \label{tab_of_cat}
    \end{table}

\section{Gate Fidelity}\label{appendix_gate_fidelity}

To quantify the accuracy of an implemented transformation, we make use of the average gate fidelity to compute the closeness of the channel $ \mathcal{E} $ to the target gate $\hat{U} $. The gate fidelity is defined as  \cite{Nielsen_2002} 
\begin{eqnarray}\label{average_fidelity}
    &&F_{\text{avg}} (\hat{U},\mathcal{E}) = \int d\psi \bra{\psi} \hat{U}^\dag\mathcal{E} (\ket{\psi}\bra{\psi})\hat{U}\ket{\psi},
\end{eqnarray}
which corresponds to the average state fidelity after being transformed by $\mathcal{E}$ and $\hat{U}$. If we express the channel in terms of Kraus operators,
\begin{eqnarray}
    \mathcal{E}(\hat{\rho}) = \sum_{i}\hat{V}_i \hat{\rho}\hat{V}^\dag_i, 
\end{eqnarray} 
the gate fidelity is given by
\begin{eqnarray}\label{fidelity_expression}
     F_{\text{avg}}(\hat{U},\mathcal{E}) = \sum_{i}\frac{\left|\text{Tr}\left(\hat{U}^\dag\hat{V}_{i}\right)\right|^2+\mathcal{D}}{\mathcal{D}(\mathcal{D}+1)},
\end{eqnarray}
where $\hat{V}_i$ are the operators and $\hat{\rho}$ is a $\mathcal{D}\times\mathcal{D}$ dimension density matrix. The gate infidelity is thus $1- F_{\text{avg}}$.

\section{Multi-tone Driving Method}\label{appendix_multi}

The conventional implementation of a SNAP gate is to drive the hybrid qumode-qubit system with multiple frequency tones \cite{SNAP_2015, PhysRevLett.133.260802}, each resonating with the energy gap between the dressed-state pair $\{\ket{\downarrow_n},\ket{\uparrow_n}\}$. In this appendix, we describe such a multi-tone driving method for both dispersive coupling and JC systems. 

\subsection{Dispersive coupling systems}

For dispersive coupling systems, each tone is chosen to resonate with the transition between the dressed-state pair $\ket{\downarrow_n} = \ket{n}\ket{g}$ and $\ket{\uparrow_n} = \ket{n}\ket{e}$, thereby inducing Rabi oscillations between the states. To engineer the SNAP gate~\eqref{U_SNAP}, the dressed state $\ket{n}\ket{g}$ is first mapped to $e^{i\Theta_n}\ket{n}\ket{e}$ via qubit driving 
\begin{eqnarray}
    \hat{H}_{\text{D}} = \sum_{n=0}^{N_{\text{max}}} \Omega_n e^{- i n \chi t} \ket{e}\bra{g}  + \text{h.c.},
\end{eqnarray}
where the Rabi frequencies are 
\begin{eqnarray}
    \Omega_n = -i \pi e^{-i\Theta_n} /2T_{\text{total}},
\end{eqnarray}
$\Theta_n$ are the target phase angles for the SNAP gate and $T_{\text{total}}$ is gate time that we choose. 

In qumode-qubit hybrid systems, any implemented transformation generally introduces the mapping
\begin{eqnarray}\label{C3}
    \ket{\psi}\ket{g}\to \sum_{i=g,e}\hat{V}_i\ket{\psi}\ket{i},
\end{eqnarray}
where $\ket{\psi} = \sum_{n=0}^\infty c_n\ket{n} $ is an arbitrary qumode state and $\hat{V}_i$ are the qumode operators. After tracing out the qubit, the resultant qumode channel is therefore
\begin{eqnarray}\label{e_impl}
    \mathcal{E}_{\text{res}}(\ket{\psi}\bra{\psi}) = \sum_{i=g,e} \hat{V}_{i}\ket{\psi}\bra{\psi}\hat{V}_{i}^\dag .
\end{eqnarray}
Then, the fidelity of a SNAP gate for the bosonic states with at most $N_{\text{max}}$ boson can be evaluated by Eq.~\eqref{fidelity_expression} with $\hat{U}=\hat{U}_{\text{SNAP}}$, $\mathcal{E}=\mathcal{E}_{\text{res}}$ and $\mathcal{D}=N_{\text{max}}+1$. 

\subsection{JC Systems}

For JC systems, the dressed states are the eigenstates of the JC interaction, i.e. Eq.~\eqref{dressed_state_JC}. To implement the SNAP gate, the computational basis states are first transformed to the dressed states, i.e. $\ket{n}\ket{e} \to \ket{\downarrow_n}$. As the computational states are superpositions of the dressed states, 
\begin{eqnarray}
    \ket{n}\ket{e}&=& -(\ket{\downarrow_n}-\ket{\uparrow_n})/\sqrt{2},\nonumber\\
    \ket{n}\ket{g}&=& (\ket{\downarrow_{n-1}}+\ket{\uparrow_{n-1}})/\sqrt{2},
\end{eqnarray}
our aim is to implement the transformation 
\begin{eqnarray}\label{transitions_JC}
    \hat{U}_{\text{hybrid}}\ket{\downarrow_n} &=& \frac{-e^{i\zeta_n/2}\ket{\downarrow_n}-e^{-i\zeta_n/2}\ket{\uparrow_n}}{\sqrt{2}},\nonumber\\
     \hat{U}_{\text{hybrid}}\ket{\uparrow_n} &=& \frac{e^{i\zeta_n/2}\ket{\downarrow_n}-e^{-i\zeta_n/2}\ket{\uparrow_n}}{\sqrt{2}},
\end{eqnarray}
where $e^{i\zeta_n/2}$ is a known phase that does not affect the subsequent phase gate engineering and can be canceled by reversing the
hybridization in the final step.

We consider a multi-tone modulation of qubit frequency,
\begin{eqnarray}
    \hat{H} &=& \sum_{n=0}^{N_{\text{max}}} \Delta_n(t) \hat{\sigma}_z \nonumber\\
    &=& \sum_{n=0}^{N_{\text{max}}} -\Delta_n(t) \Big(\ket{0}\bra{0}\otimes\ket{g}\bra{g}\nonumber\\
    &+&\sum_{m=0}^\infty\left(\ket{\downarrow_m}\bra{\uparrow_m}+\text{h.c.} \right)\Big),\label{drive_JC2}
\end{eqnarray}
where
\begin{eqnarray}
    \Delta_n(t) = \Delta_0 \sin \left(2\lambda\sqrt{n+1} t + \varphi_n\right),
\end{eqnarray}
and $\Delta_0$ characterizes the amplitude of the modulation.
Except that the ground state $\ket{0}\ket{g}$ remains unchanged, each frequency tone resonates with the energy gap between $\ket{\downarrow_n} $ and $ \ket{\uparrow_n}$ and induces Rabi oscillation between the states. Specifically, we choose $\varphi_n=0$ and $\Delta_0 = \pi/(2T_f)$ for realizing the hybridization, where $T_f$ denotes our choice of gate time for the hybridization process. 

To evaluate the fidelity of the hybridization for $N_{\text{max}}+1$ bosonic levels, we use Eq.~\eqref{fidelity_expression} with $\hat{U}= \hat{U}_{\text{hybrid}}$, $\mathcal{E}(\hat{\rho}) = \hat{U}_{\text{impl}}\hat{\rho}\hat{U}_{\text{impl}}^\dag $, where $\hat{U}_{\text{impl}}$ is the implemented qumode-qubit unitary operator. Because the hybridization acts on all relevant dressed states, the operator dimension is doubled, i.e. $\mathcal{D} = 2(N_{\text{max}}+1)$. The results are plotted in Fig.~\ref{fig_JC} (a).

Next, we consider the phase modification process,
$\ket{\downarrow_n} \to e^{i\Theta_n}\ket{\downarrow_n} $. It can be implemented with the Hamiltonian in Eq.~\eqref{drive_JC2} with the parameters
\begin{eqnarray}
    \varphi_n= \begin{cases} 
    \sqrt{n+1}T_f/2-\Theta_n/2 \\ \quad \quad \quad  \quad  \text{for} \quad 0<t<T_f/2 \\
    \pi-\sqrt{n+1}T_f/2+ \Theta_n/2 \\ \quad \quad  \quad \quad\text{for} \quad T_f/2<t<T_f
    \end{cases},
\end{eqnarray}
and the modulation strength $\Delta_0=2\pi/T_f$. 
For both multi-tone and QSP methods, the implemented unitary generally transforms the dressed states as
\begin{eqnarray}
    \ket{\downarrow_n} \to  V_{e,n}\ket{\downarrow_n} + V_{g,n}\ket{\uparrow_n},
\end{eqnarray}
where $V_{e(g),n}$ are complex amplitudes. Ideally, $V_{e,n}=e^{i\Theta_n}$ and $V_{g,n}=0$.
To quantify the fidelity of this process, we assume that the hybridization operations~\eqref{transitions_JC} are perfect. The resulting transformation can thus be expressed as
\begin{eqnarray}
   && \ket{n}\ket{e}\to  e^{i\zeta_n/2}\ket{\downarrow_n} \nonumber\\
    &&\to e^{i\zeta_n/2}(V_{e,n}\ket{\downarrow_n} + V_{g,n}\ket{\uparrow_n})\nonumber\\
    &&\to V_{e,n}\ket{n}\ket{e} + V_{g,n}\ket{n}\ket{g}.
\end{eqnarray}
After tracing out the qubit, the resulting channel is given by Eq.~\eqref{e_impl} with $\hat{V}_{i} = \sum_{n=0}^{N_{\text{max}}}V_{i,n}\ket{n}\bra{n}$. Thus, the fidelity of a phase modification for the bosonic states with at most $N_{\text{max}}$ boson is evaluated via Eq.~\eqref{fidelity_expression} with $\hat{U}=\hat{U}_{\text{SNAP}}$, $\hat{V}_{i}$ defined above, and $\mathcal{D}=N_{\text{max}}+1$. 
The results are plotted in Fig.~\ref{fig_JC} (b).

\section{Kernel for the JC Systems}\label{appendix_JC}

In this appendix, we present our procedure for determining the integer powers $s$ and $h$ used in the kernel method for JC systems, Eqs.~(\ref{kernel_modified}) and (\ref{kernel_modified_I}). 
We construct the polynomial $F(z)$ to take the following form:
\begin{eqnarray}
    F(z) &=& \sum_{n=0}^{N_{\text{max}}}\Big[ \text{Re}\left[A_n \right]\mathcal{K}^R_n(z) + \text{Im}\left[A_n\right] \mathcal{K}^I_n(z)\Big],\nonumber\\
\end{eqnarray}
where $A_n$ are arbitrary complex amplitudes that are bounded $|A_n|\leq 1$.

We begin by considering the norm of the polynomial $F(z)$ and applying the triangle inequality to establish an upper bound:
\begin{eqnarray}
    |F(z)| &=& \left| \sum_{n=0}^{N_{\text{max}}} [\text{Re}\left[A_n \right] \mathcal{K}_n^R(z) + i\text{Im}\left[A_n \right] \mathcal{K}_n^I(z)]\right| \nonumber\\
     &\leq& \sum_{n=0}^{N_{\text{max}}} \left| \text{Re}\left[A_n \right] \mathcal{K}_n^R(z) + i\text{Im}\left[A_n \right] \mathcal{K}_n^I(z)\right|.\nonumber\\
    \end{eqnarray}
Then, we further bound this using the triangle inequality of complex numbers and the bound of $A_n$:
\begin{eqnarray}
       |F(z)| &\leq& \sum_{n=0}^{N_{\text{max}}} \left[ \left| \text{Re}\left[A_n \right]\mathcal{K}_n^R(\Phi)\right| + \left|\text{Im}\left[A_n \right] \mathcal{K}_n^I(\Phi)\right|\right]\nonumber\\
    &\leq& \sum_{n=0}^{N_{\text{max}}} \left[ \left|  \mathcal{K}_n^R(\Phi)\right| + \left|\mathcal{K}_n^I(\Phi)\right|\right].
\end{eqnarray}

Because this bound is independent of the target amplitude, we determine the parameters $s$ and $h$ by ensuring 
\begin{eqnarray}
    \sum_{n=0}^{N_{\text{max}}} \left[ \left|  \mathcal{K}_n^R(\Phi)\right| + \left|\mathcal{K}_n^I(\Phi)\right|\right] \leq 1, \label{criterion}
\end{eqnarray}
for all $\Phi$. 
Direct numerical calculations identify several sets of parameters that satisfy this criterion. Since our objective is to minimize the total implementation time~\eqref{JC_total_time},
we focus on the parameter sets that minimize the numerator. Table~\ref{tab_of_parameters} summarizes some of these numerically optimized parameters. While the values of $s$ and $h$ presented here guarantee that the polynomial must be bounded by unity, smaller values might be feasible for some sets of target phases.

 \begin{table}[h!]
    \centering
    \begin{tabularx}{0.45\textwidth}{|>{\centering\arraybackslash}X|>{\centering\arraybackslash}X|>{\centering\arraybackslash}X|>{\centering\arraybackslash}X|>{\centering\arraybackslash}X|>{\centering\arraybackslash}X|}
    \hline
      $N_{\text{max}}$   &  2 & 3 & 4 & 5 & 6 \\
      \hline
        $s$ &  2 & 2 & 2 & 3 & 3  \\
        \hline
        $h$ &  12 & 34 & 44 & 42 & 60 \\
       \hline
    \end{tabularx}
    \caption{Table of the parameters $s$ and $h$ obtained from the criterion~\eqref{criterion}.}
        \label{tab_of_parameters}
    \end{table}

\section{Dressed Basis for Anti-JC Systems}\label{appendix_aJC}

As the counterpart of JC interaction, anti-JC interaction is also commonly realized in the first blue-sideband regime \cite{RevModPhys.75.281} and cavity QED systems \cite{RevModPhys.87.1379, PhysRevLett.90.027903}. Unlike the JC interaction, anti-JC interaction does not conserve the number of excitations, since it creates both a qubit and a bosonic excitation simultaneously and vice versa. Nevertheless, its dressed states are similar to those of JC systems. 
In this appendix, we will show how to implement QSP in anti-JC systems.

The Hamiltonian of an anti-JC system is given by
\begin{eqnarray}\label{Ham_AJC}
    \hat{H}_{\text{AJC}} = \lambda_A \left(\hat{a}^\dag \ket{e}\bra{g} + \hat{a} \ket{g}\bra{e}\right),
\end{eqnarray}
where $\lambda_A$ is the anti-JC interaction strength. In this system, the dressed states are the eigenstates of Eq.~\eqref{Ham_AJC}:
\begin{eqnarray}
    \ket{\downarrow_n} = \frac{1}{\sqrt{2}}\left(\ket{n+1}\ket{e} - \ket{n}\ket{g} \right),\\
    \ket{\uparrow_n} = \frac{1}{\sqrt{2}}\left(\ket{n+1}\ket{e} + \ket{n}\ket{g} \right),
\end{eqnarray}
for integer $n\in [0,\infty)$. In fact, JC and anti-JC dressed states are related by interchanging $\ket{g} \leftrightarrow\ket{e}$. Therefore, for anti-JC systems, the hybrid ground state $\ket{0}\ket{e}$ will not form any dressed pair, and the evolution in the dressed basis, $\{\ket{\downarrow_n}\equiv(1,0)^T, \ket{\uparrow_n}\equiv (0,1)^T\}$, is given by,
\begin{eqnarray}
\begin{pmatrix}
e^{i\Phi_n/2} & 0 \\
0 & e^{-i\Phi_n/2}
\end{pmatrix} =
e^{-i\Phi_n/2} \hat{Z}_{\Phi_n}.
\end{eqnarray}
Here, the phase-shift angle is given by $\Phi_n = \lambda_A T \sqrt{n+1}$.

Like JC systems, in anti-JC systems, only the qubit $z$-axis rotation can retain the dressed states within the same subspace. This operation can be induced by the detuning $\Delta \hat{\sigma}_z$ and it introduces only an $x$-axis rotation in the dressed-state basis, i.e. a Rabi oscillation between $\ket{\uparrow_n}$ and $\ket{\downarrow_n}$. Since only single-axis rotations can be implemented, oQSP should be considered for anti-JC.

Since JC and anti-JC systems share the same mathematical structure, the same procedure and kernel method can be applied to engineer operations with the replacement $\ket{g} \to \ket{e}$, $\ket{e} \to \ket{g}$ and $\lambda\to \lambda_A$. 

\bibliography{aps_reference}
\pagestyle{plain}

\end{document}